\documentclass[aps,prc,preprint,groupedaddress,12pt]{revtex4-1}
\usepackage{natbib}
\usepackage{amsmath}
\usepackage{graphicx}
\usepackage{hyperref}
\usepackage{multirow}
\usepackage{bm}

\begin{document}

\preprint{JLAB-THY-14-1981}


\title{Momentum distributions for $^2$H$(e,e'p)$}

\author{William P. Ford$^{(1)}$}
\email[]{wpford@jlab.org}
\author{Sabine Jeschonnek$^{(2)}$}
\email[]{sjeschonnek@lima.ohio-state.edu}
\author{J. W. Van Orden$^{(3,4)}$}
\email[]{vanorden@jlab.org}
\affiliation{\small \sl 
(1) Department of Physics and Astronomy, University of Southern Mississippi, Hattiesburg, MS 39406 \\ 
(2) The Ohio State University, Physics Department, Lima, OH 45804 \\ 
(3) Department of Physics, Old Dominion University, Norfolk, VA 23529 \\ 
and\\ (4) Jefferson Lab\footnote{Notice: Authored by Jefferson Science Associates, LLC under U.S. DOE Contract No. DE-AC05-06OR23177.
The U.S. Government retains a non-exclusive, paid-up, irrevocable, world-wide license to publish or reproduce this manuscript for U.S. Government purposes},
12000 Jefferson Avenue, Newport News, VA 23606}
\date{\today}
\begin{abstract}
\begin{description}
\item[Background] A primary goal of deuteron electrodisintegration is the possibility of extracting the deuteron momentum distribution.
This extraction is inherently fraught with difficulty, as the momentum distribution is not an observable and the extraction relies
on theoretical models dependent on other models as input.
\item[Purpose] We present a new method for extracting the momentum distribution which takes into account a wide variety of model inputs thus providing a theoretical uncertainty due to the various model constituents.
\item[Method] The calculations presented here are using a Bethe-Salpeter like formalism with a wide variety of bound state wave functions, form factors, and final state interactions. We present a method to extract the momentum distributions from experimental cross sections, which takes into account the theoretical uncertainty from the various model constituents entering the calculation. 
\item[Results] In order to test the extraction pseudo-data was generated, and the extracted ``experimental'' distribution, which has theoretical uncertainty from the various model inputs, was compared with the theoretical distribution used to generate the pseudo-data. 
\item[Conclusions] In the examples we compared, the original distribution was typically within the error band of the extracted distribution. The input wave functions do contain some outliers which are discussed in the text, but at least this procedure can provide an upper bound on the deuteron momentum distribution. Due to the reliance on the theoretical calculation to obtain this quantity any extraction method should account for the theoretical error inherent in these calculations due to model inputs.
\end{description}
\end{abstract}

\maketitle

\section{Introduction}\label{sec:Introduction}


One of the primary reasons for measuring deuteron electrodisintegration at large missing momenta is the possibility of finding small 
(exotic) 
configurations of quarks which are of small size and could possibly be examined by determining the deuteron momentum distribution at large missing momenta. This requires that the momentum distribution be extracted from the experimental cross sections. This is in general not possible since the cross section is obtained from squares of the transition matrix element denoted by $<p_1,s_1;p_2,s_2;(-)|J^\mu_{\rm em}|P,\lambda_d>$ where $|P,\lambda_d>$ is the state of the initial deuteron with total momentum $P$ and helicity $\lambda_d$, $<p_1,s_1;p_2,s_2;(-)|$ is the proton-neutron scattering state with incoming wave boundary conditions and $J^\mu_{\rm em}$ is the electromagnetic current operator. The relationship between the interactions producing the initial and final states and the electromagnetic current operator is constrained by the requirement of electromagnetic current conservation, which may appear as a commutation relation between the hamiltonian and the components of the current operator or in the case of Bethe-Salpeter based formulations, such as the one used in this work, as two-body Ward identities\cite{Adam:1997cx}. As a result, construction of a consistent description of the matrix element will result in different partitions into initial and final states, and the current operator which depends on the basic formalism used to model the matrix element. This implies that the momentum distribution of the initial deuteron is model dependent\cite{Furnstahl:2010wd} and can only be determined approximately if there are sound theoretical grounds for ignoring final state interaction and two-body electromagnetic currents.

At relatively small momentum transfers the interactions and currents can be constructed consistently by means of chiral perturbation theory, by traditional nonrelativistic potential models with some input from meson exchange models or in terms of Bethe-Salpeter-like models based on meson exchange.  All of these models are constrained by fitting $np$ scattering to cross sections for energies up to slightly above pion production threshold. At present there are no consistent calculations of matrix elements at the larger momentum transfers needed to explore large missing momenta.

At large missing momenta it is therefore necessary to construct models of the matrix elements and cross section based on a set of reasonable choices for initial and final states as well as the electromagnetic current operator. This means that the available models do not conserve current and that a large number of different theoretical models are available based on the number of possible reasonable choices that are available for initial wave functions or their equivalent, for final state interactions and for the current operators, as well as differences due to alternate theoretical choices used to produce the matrix elements.

The basic experimental approach to extracting an approximate momentum distribution is to search for kinematic regions where the effects of final-state interactions and 2-body currents are small\cite{Boeglin_highmom}. This requires input from theory that may result in a certain amount of model dependence based on the range of models that are used to select these regions. The cross sections measured for the chosen kinematics are then divided by some kinematical factors related to the deuteron cross section and a prescription for an off-shell $ep$ cross section. This results in a reduced cross section which is assumed to be close in size and shape to the deuteron ground-state momentum distribution.

The objective of this work is to examine this procedure for extracting the deuteron momentum distribution by means of producing a large number of model calculations using reasonable choices for initial state wave functions, final state interactions and nucleon electromagnetic form factors based on the Bethe-Salpeter-like approach of \cite{JVO_2008_newcalc,JVO_2009_tar_pol,JVO_2009_ejec_pol,FJVO}. This allows us to study the properties of the usual procedure and to generate a statistical treatment of theoretical corrections which can be used to improve the description of the momentum distribution along with a theoretical error band.  In doing this we choose the kinematics of the approved Jefferson lab experiment E1210003\cite{Boeglin:2014aca}. Similar calculations could be made using different frameworks\cite{CiofidegliAtti:2004jg,Laget:2004sm,Sargsian:2009hf} and would in combination with those presented here help to establish the possible variations in theoretical models.

This paper is organized as follows: In Section \ref{sec:Theoretical_Framework} , we lay out the theoretical framework for our calculations. In Section \ref{sec:Model_Dependencies} we discuss our choices of wave functions, electromagnetic form factors and final state interaction models that we use in this work. Finally in Section \ref{sec:Calculations} we discuss the characteristics of the model calculations which are produced. The method that we propose to provide theoretical corrections and error to the reduced cross sections to obtain improved momentum distributions is presented in Section \ref{sec:NewMethod}. Section \ref{sec:Results} presents several tests of this method obtained by using a selection of model calculations as pseudo-data and comparing to the actual momentum distributions associated with the model. A summary of this work and conclusions drawn from it are contained in \ref{sec:Summary}.

\section{Theoretical Framework} \label{sec:Theoretical_Framework}

\begin{figure}
\centerline{\includegraphics[height=2in]{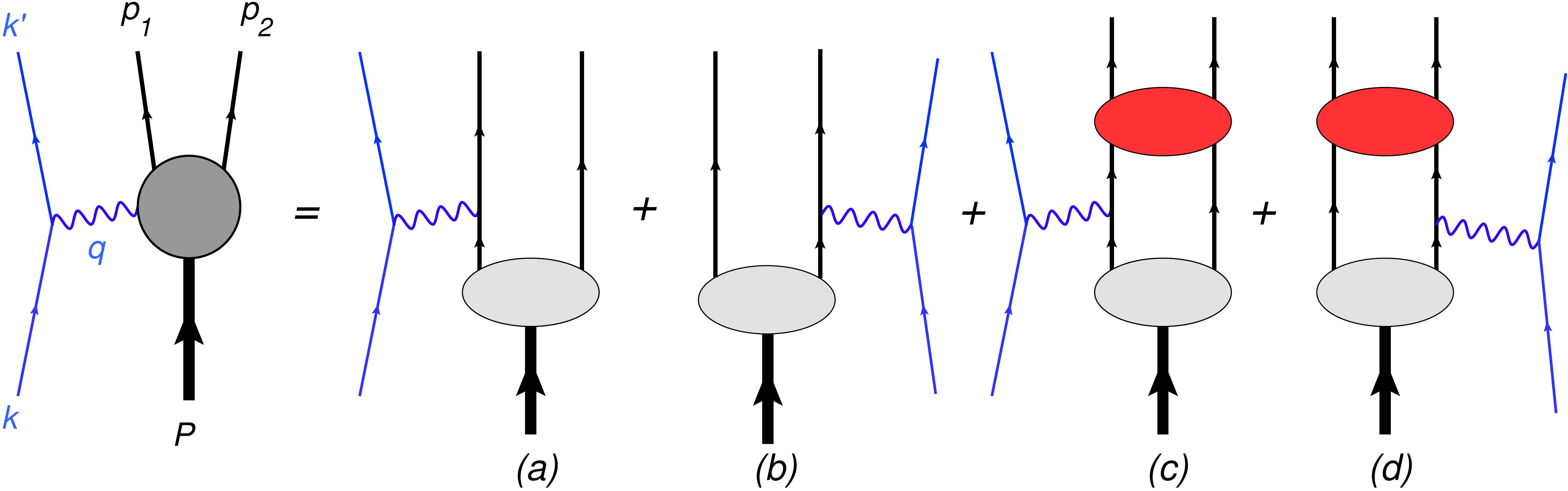}}
\caption{(color online) Feynman diagrams representing the impulse approximation to deuteron electrodisintegration. In all diagrams particle 1 is a proton and particle 2 is a neutron. 
}
\label{fig:IA_diagrams}
\end{figure}

The calculations used in this work use the formalism of \cite{JVO_2008_newcalc,JVO_2009_tar_pol,JVO_2009_ejec_pol,FJVO} which is based on approximations to the Bethe-Salpeter equation. For large $Q^2$ it is not possible at this point to construct a consistent meson exchange model for the complete matrix elements for deuteron electrodisintegration. For this reason the calculations are performed using bound states, current operators and final state interactions from a variety of sources that will result in a violation of current conservation. The current consensus is that two-body currents give no substantial contribution at large $Q^2$ and that they can safely be ignored. The calculations are therefore performed in impulse approximation as represented by Fig \ref{fig:IA_diagrams}. 
Diagram \ref{fig:IA_diagrams} (a) represents the plane wave impulse approximation (PWIA). Diagrams \ref{fig:IA_diagrams} (a) plus \ref{fig:IA_diagrams} (b) represent the plane wave born approximation (PWBA). The t matrices providing the final state interactions (FSI) in diagrams \ref{fig:IA_diagrams} (c) and \ref{fig:IA_diagrams} (d) are properly antisymetrized assuming that isospin is a good quantum number. In diagrams \ref{fig:IA_diagrams}(a)-\ref{fig:IA_diagrams}(d) the initial bound state is represented by the spectator equation deuteron vertex function (\ref{eq:vertex_function}).

The unpolarized cross section for deuteron electrodisintegration can be written as
\begin{equation}
\frac{ d \sigma^5}{d \epsilon' d \Omega_e d \Omega_p} =  \frac{m_p \, m_n \, p_p}{8 \pi^3 \, M_d} \,
\sigma_{Mott} \,
f_{rec}^{-1} \,
 \left[  v_L R_L +   v_T R_T
 + v_{TT} R_{TT}\cos 2\phi_p + v_{LT}R_{LT}\cos\phi_p
\right] 
\label{def5sigma}\,,
\end{equation}
where the Mott cross section is
\begin{equation}
\sigma_{Mott} = \left ( \frac{ \alpha \cos(\theta_e/2)} {2
\varepsilon \sin ^2(\theta_e/2)} \right )^2
\end{equation}
and the recoil factor is
\begin{equation}
f_{rec} = \left| 1+ \frac{\omega p_p - E_p q \cos \theta_p} {M_d \, p_p}
\right| \, . \label{defrecoil}\,.
\end{equation}
The $v_i$ are kinematical factors defined as
\begin{eqnarray}
v_L&=&\frac{Q^4}{q^4}\\
v_T&=&\frac{Q^2}{2q^2}+\tan^2\frac{\theta_e}{2}\\
v_{TT}&=&-\frac{Q^2}{2q^2}\\
v_{LT}&=&-\frac{Q^2}{\sqrt{2}q^2}\sqrt{\frac{Q^2}{q^2}+\tan^2\frac{\theta_e}{2}}
\end{eqnarray}
If the response tensor is defined as
\begin{equation}
W^{\mu\nu}=\frac{1}{3}\sum_{s_1,s_2,\lambda_d}\left<\bm{p}_1s_1;\bm{p}_2s_2\right|J^\mu\left|\bm{P}\lambda_d\right>^* \left<\bm{p}_1s_1;\bm{p}_2s_2\right|J^\nu\left|\bm{P}\lambda_d\right>
\end{equation}
the response functions $R_K$ are defined by
\begin{eqnarray}
R_L & \equiv & W^{00} \nonumber \\
R_T & \equiv & W^{11}+W^{22} \nonumber  \\
R_{TT}\cos 2\phi_p & \equiv &  W^{22}-W^{11} \nonumber  \\
R_{LT}\cos\phi_p & \equiv & 2 \sqrt{2}\, \Re(W^{01})  \, , \label{defresp}
\end{eqnarray}

For convenience we define
\begin{equation}
\sigma_{eD}\equiv\frac{ d \sigma^5}{d \epsilon' d \Omega_e d \Omega_p}
\end{equation}
It is conventional to define a reduced cross section as
\begin{equation}
\sigma_{red}=\frac{\sigma_{eD}}{k \sigma_{ep}}\,,
\end{equation}
where $\sigma_{ep}$ is an off shell electron proton cross section usually chosen to be either deForrest cc1 or cc2\cite{DeForest:1983vc} and $k$ is some appropriate combination of factors obtained to reproduce the deuteron electrodisintegration cross section under the assumption that the PWIA cross section factorizes. A demonstration of how such a factorization can be obtained from the formalism used here is contained in Appendix \ref{app:factor}.

\section{Model Constituents}\label{sec:Model_Dependencies}
In extracting the momentum distributions one must rely on accurate theoretical models.
The primary objective of this work is to examine the variation in calculated cross sections on a variety of reasonable choices for the constituents ,
and thereby provide the theoretical uncertainty that can be expected when extracting the approximate momentum distributions.

The three major uncertainties that can influence the calculation stem from form factors, 
the deuteron wave function, and final state interactions.
Our approach is to perform our calculation using as many 
possible variations of each of these in order to understand the way each can influence the calculation.
The final result is represented as the mean and the standard deviation is treated as the 
theoretical uncertainty due to input model dependencies.
The various models we use as input are given in Table \ref{tab:model_inputs}.

\begin{table}\caption{Model inputs to the calculation.}
 \begin{center}
\begin{tabular}{ c | c | c  } 
\hline
Final State Interactions  & Form Factors & Deuteron Wave Function \\
\hline \hline
\multirow{8}{4em}{Regge \cite{FVO_Reggemodel,FJVO,Ford:2013wxa} \\SAID \cite{SAIDdata,Arndt07,Arndt00} } & \multirow{8}{4em}{GKex05\cite{FF_GK_05_2_Lomon_02,FF_GK05_1_Lomon_06} \\ AMT\cite{FF_AMT_07} \\ MMD\cite{FF_MMD1_Mergell_96}} & IIB \cite{Gross:1991pm}\\ 
			 &  & WJC 1\cite{Gross:2007jj} \\ 
                         &  & WJC 2\cite{Gross:2007jj}  \\ 
                         &  & AV18 \cite{Wiringa95} \\ 
			&  & CD Bonn \cite{Machleidt:2000ge}\\ 
                               &  & NIMJ 1 \cite{Stoks:1994wp}\\ 
                                &  & NIMJ 2\cite{Stoks:1994wp} \\ 
                                &  & NIMJ 3\cite{Stoks:1994wp} \\ 
\hline
\end{tabular}
\end{center} \label{tab:model_inputs}
\end{table}

 All of these form factors and wave functions are widely used in the literature. Clearly,
they introduce deviations in the calculations, and these deviations vary in size from tiny to significant, 
depending on the kinematics. 

\subsection{Wave Functions}

In the calculations performed here we use eight different wave functions. Those labeled IIB\cite{Gross:1991pm}, WJC 1 and WJC 2\cite{Gross:2007jj} are the results of fitting the spectator or Gross equation to $NN$ scattering data. WJC 1 and WJC 2 are associated with fits with $\chi^2$ per degree of freedom of approximately 1. IIB is the result of earlier fits with larger $\chi^2$ but was used successfully in calculating electron-deuteron elastic scattering and has a momentum distribution comparable to that of the majority of non-relativistic potentials and is used here to provide continuity with earlier results. These are technically not wave functions but are the result of calculating the spectator equation vertex functions. The remaining wave functions are the nonrelativistic potentials AV18\cite{Wiringa95}, CD Bonn\cite{Machleidt:2000ge}, NIJM 1, NIJM 2 and NIJM 3\cite{Stoks:1994wp}. All of these potentials produce fits to the $NN$ data with $\chi^2$ per degree of freedom of approximately 1. As a result, all of the wave functions but IIB produce on-shell equivalent scattering amplitudes but differ off shell. The nonrelativistic wave functions are used in the calculations presented here by replacing $u$ and $w$ in (\ref{eqn:g1to4}) the s- and d-state wave functions for the nonrelativistic wave functions and setting $v_s$ and $v_t$ in (\ref{eqn:g1to4}) to zero. As can be seen from Appendix \ref{app:factor}, this results in the commonly used factorization of the PWIA.

\begin{figure}
{\includegraphics[height=3in]{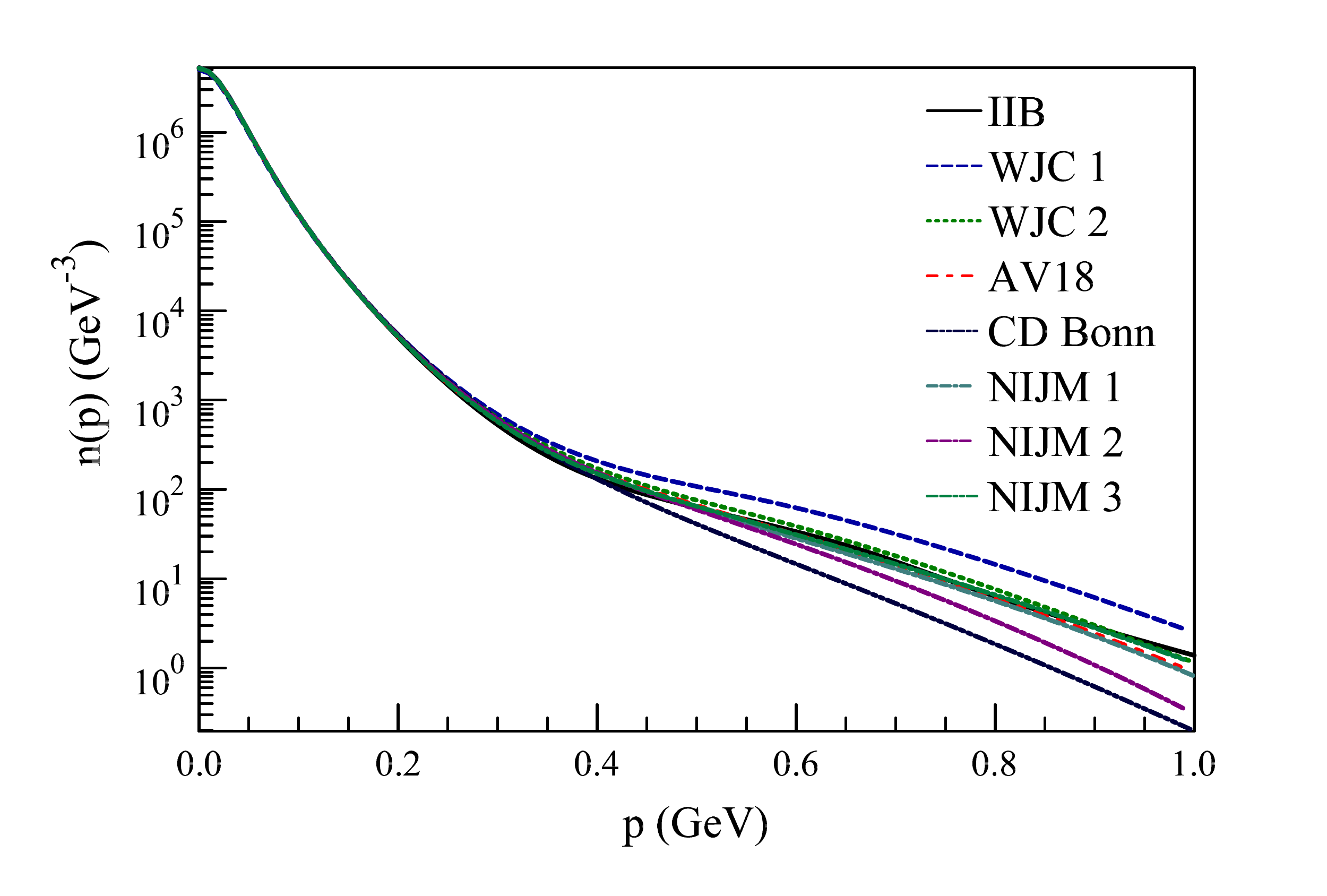}}
\caption{(color online)Momentum density distributions for the eight wave functions used in the calculations presented in this work.
}
\label{fig:mom_dist}
\end{figure}

The momentum distributions for the eight sets of initial states are shown in Fig. \ref{fig:mom_dist} using the normalization given by (\ref{eq:normalize}).
From Fig. \ref{fig:mom_dist} it can be seen that the momentum distribution for CD Bonn is the softest (has the smallest high-momentum tail) and the next softest is NIJM 2. The hardest distribution is for WJC 1. This wave function has the largest relativistic p-wave contributions that result from the presence of negative-energy projections in the spectator equation. These negative-energy projections provide a repulsive contribution to the NN force resulting in a stronger repulsive core and thus a larger high-momentum tail. The remaining wave functions provide momentum distributions which fall within a relatively narrow band.

\subsection{Form Factor Parameterizations}

\begin{figure}
{\includegraphics[height=2in]{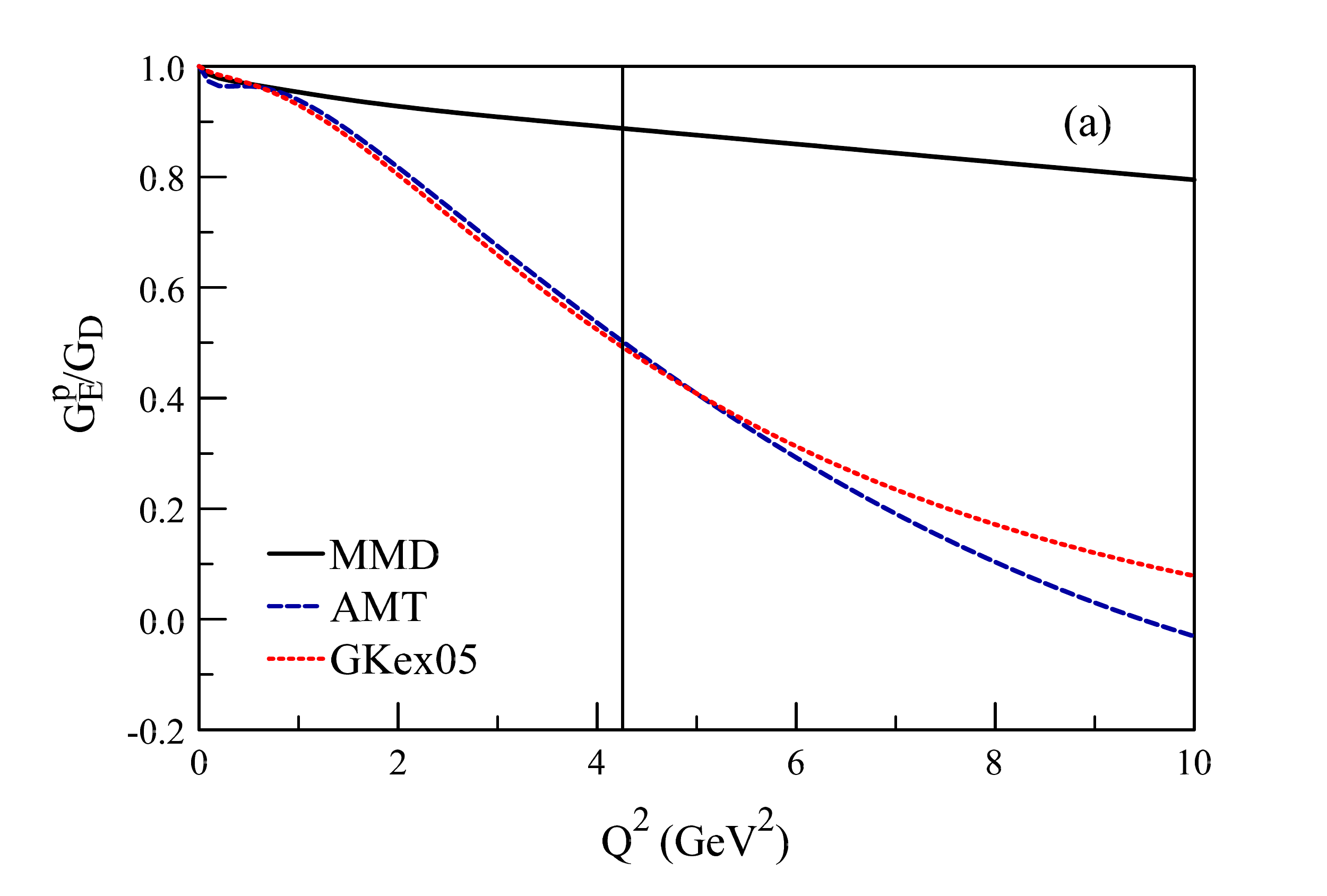}}
{\includegraphics[height=2in]{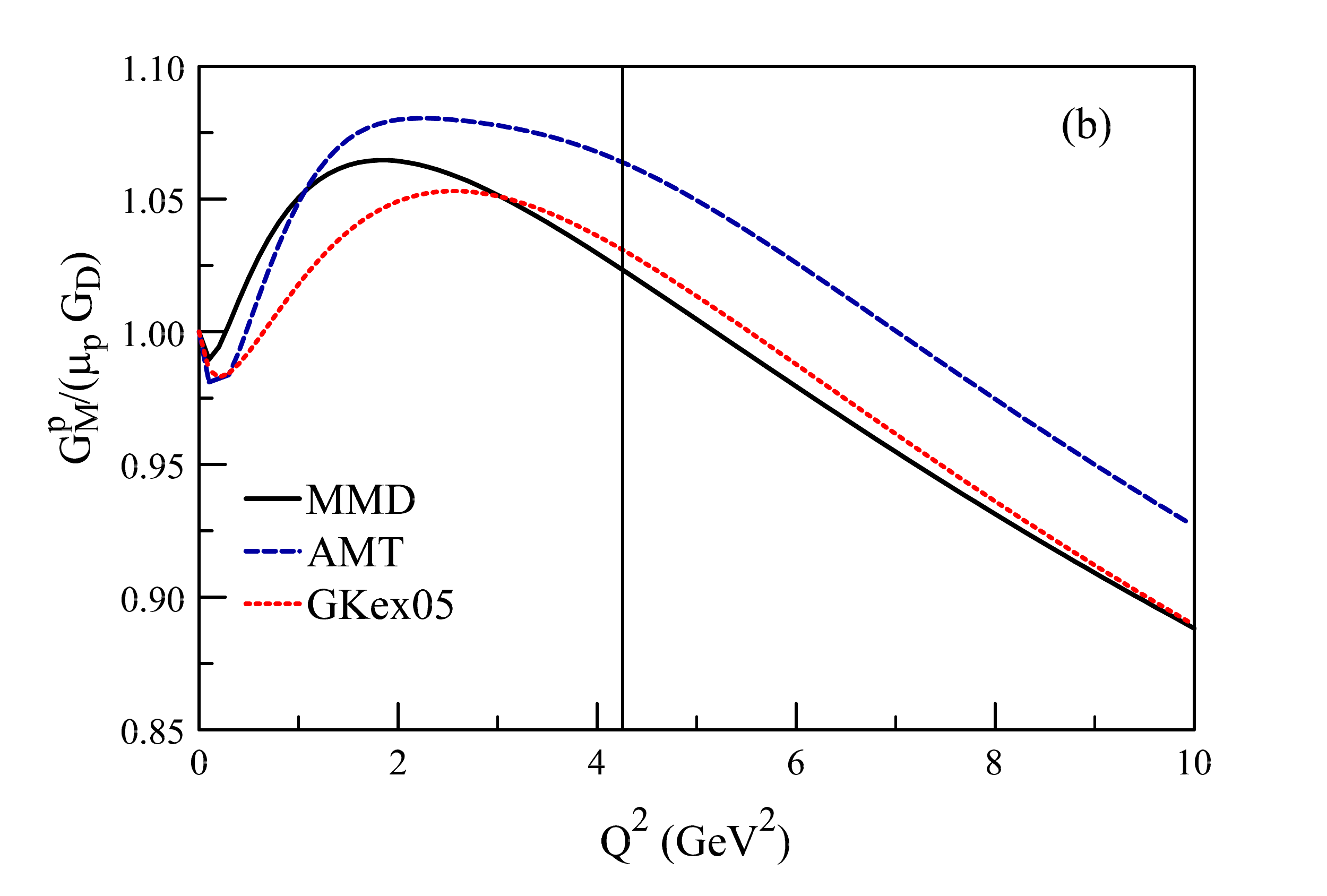}}
{\includegraphics[height=2in]{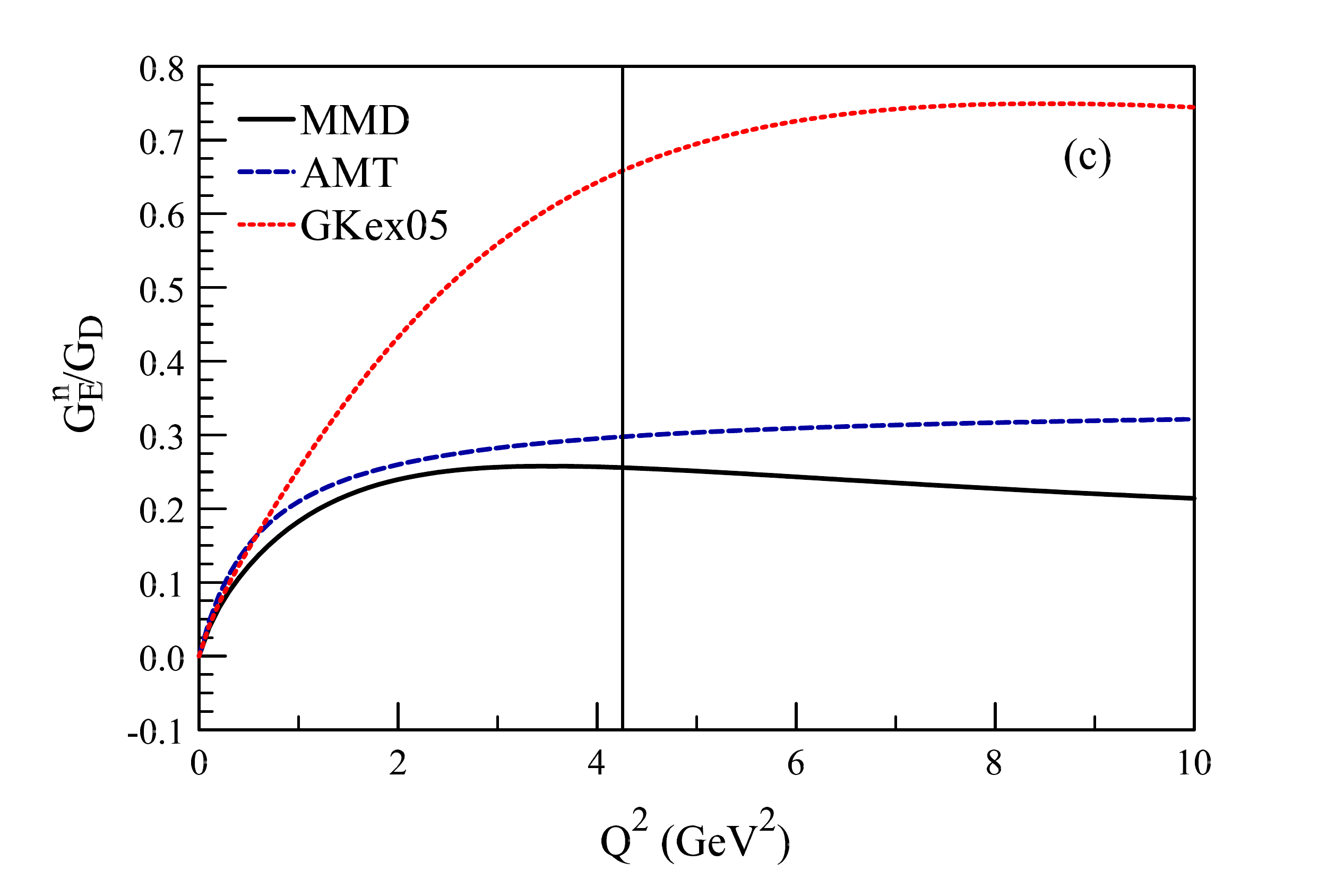}}
{\includegraphics[height=2in]{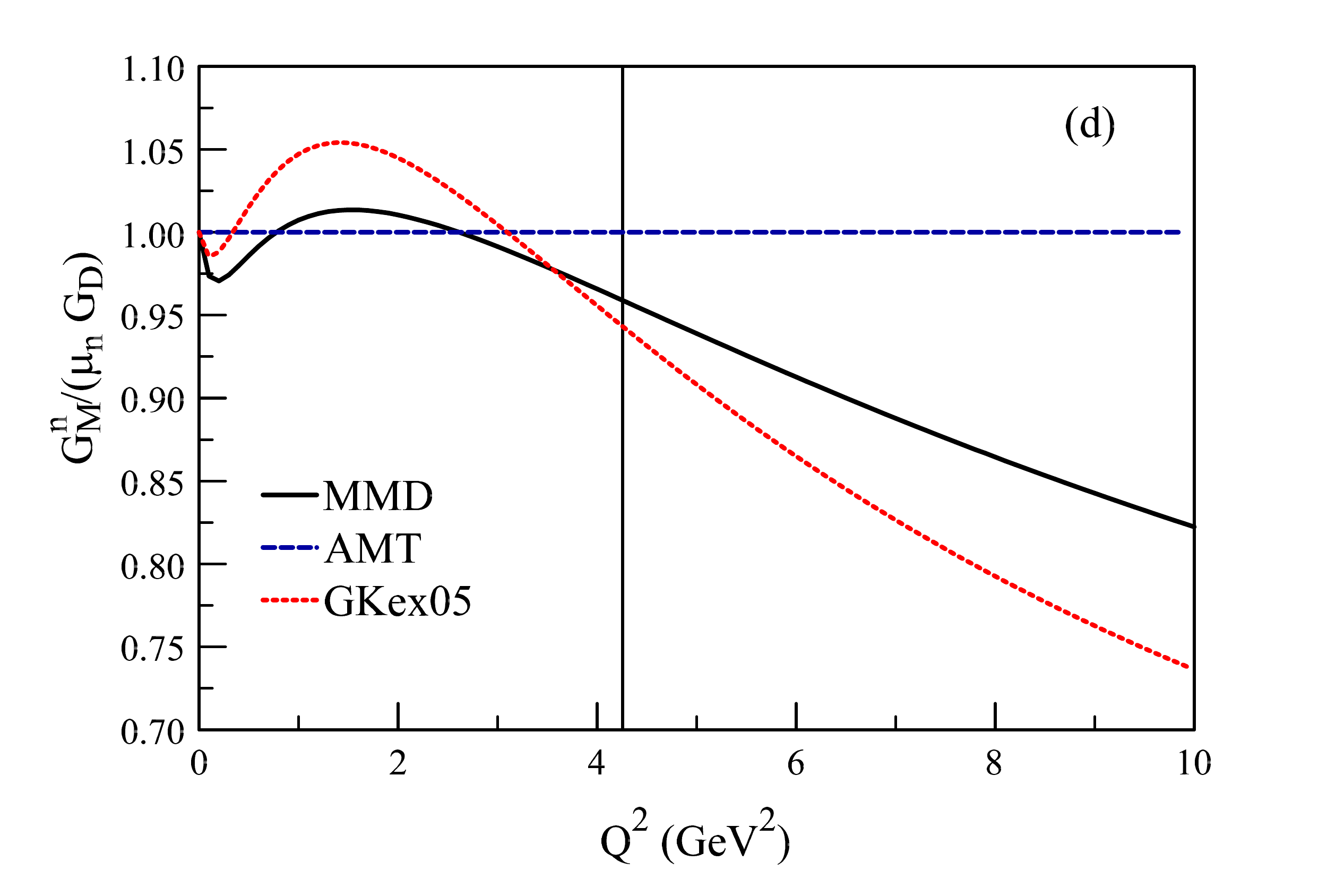}}
\caption{(color online)A comparison of three different parameterizations of the electric and magnetic form factors of the proton and neutron, divided by the corresponding dipole form factors.
dipole form factor. In each case the vertical line corresponds to $Q^2=4.25\ {\rm GeV^2}$ which is the value for the chosen kinematics.
}
\label{fig:gepfig}
\end{figure}

We use the standard Dirac-plus-Pauli form of the single nucleon current operator
\begin{equation}
\Gamma^\mu(q)=F_1(Q^2)\gamma^\mu+\frac{F_2(Q^2)}{2m}i\sigma^{\mu\nu}q_\nu\,.
\end{equation}
in the calculations presented here. We choose three different parameterizations of the form factors. 
The form factors GKex05 are the result of a vector meson dominance model (VMD) to the nucleon electromagnetic form factor data including the rapidly falling $G_E^p$ data obtained from electron-proton scattering with either polarized initial of final states. The form factors AMT are a fit to the new proton scattering data only with the usual Galster parameterization of the neutron form factors. The form factors MMD and VMD model fit to the form factor data prior to the availability of the data from polarized protons. This is included for continuity with earlier calculations and to provide a sense of the importance of the new parameterizations of $G_E^p$ at the kinematics chosen for the calculations presented here.
Figure \ref{fig:gepfig} shows the Sachs form factors divided by the equivalent simple dipole forms for $0<Q^2<10\ {\rm GeV^2}$ for the three chosen parameterizations.

\begin{figure}
\centerline
{\includegraphics[height=3in]{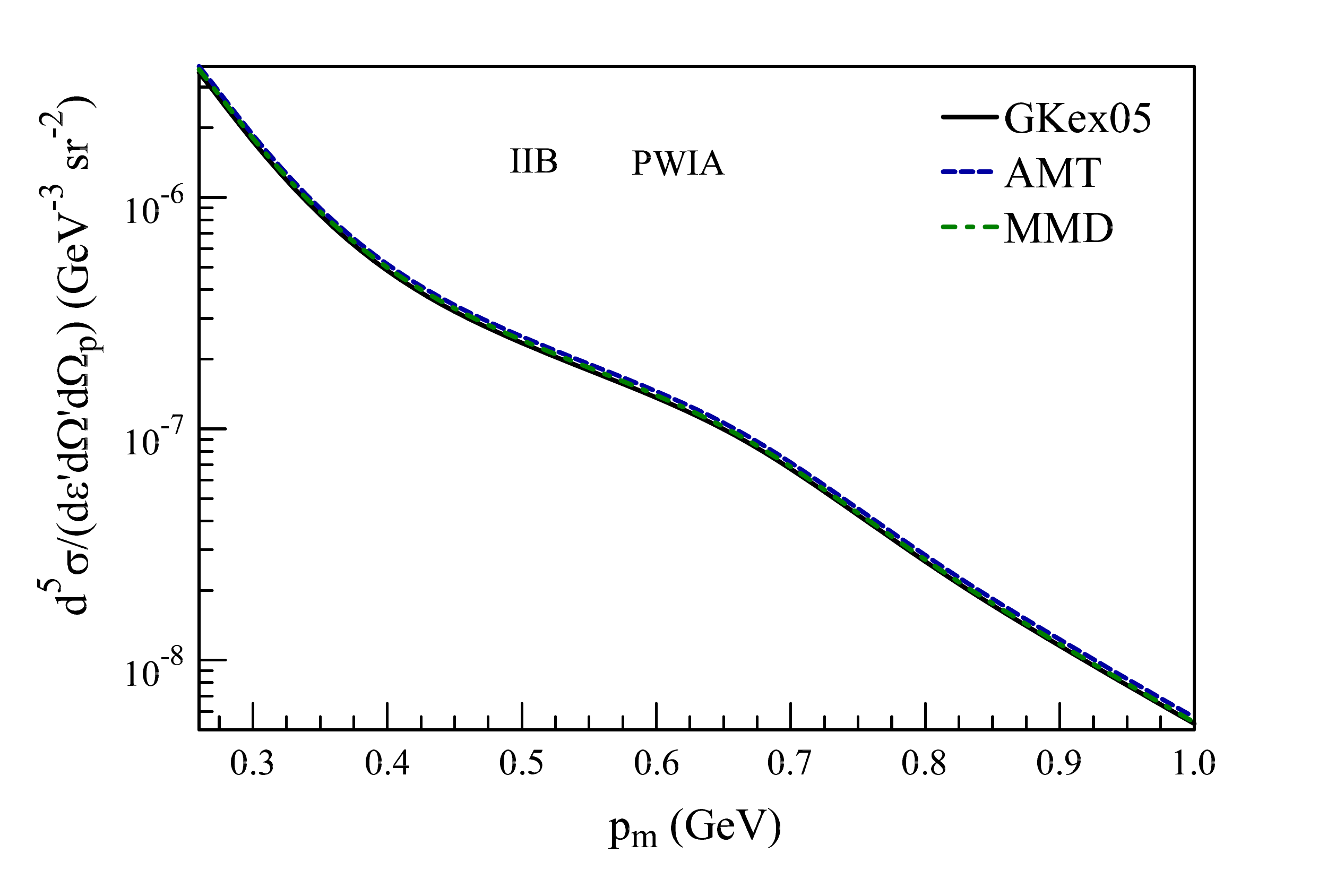}}
\caption{(color online)PWIA cross sections calculated for the E121003 kinematics with the IIB wave functions and the three form factor parameterizations. Cross sections are plotted as a function of the missing momentum $p_m$. 
}
\label{fig:pwiaff}
\end{figure}

We choose the kinematics of experiment E1210003, which is approved for running in Hall C at Jefferson Lab. These are specified by $\epsilon=12\ {\rm GeV}$, $Q^2=4.25\ {\rm GeV^2}$, $x=1.35$ and $\phi_p=180^\circ$. Figure \ref{fig:pwiaff} shows the PWIA cross section calculated at these kinematics using the IIB wave functions and the three parameterizations of the electromagnetic form factors. Although Fig. \ref{fig:gepfig} shows that the different parameterizations vary considerably at this point, Fig. \ref{fig:pwiaff} shows that variation in the PWIA cross section due to the form factors is relatively small but non-negligible. Note however that the PWIA uses only the proton form factors.

\subsection{Final State Interactions}

\begin{figure}
{\includegraphics[height=3in]{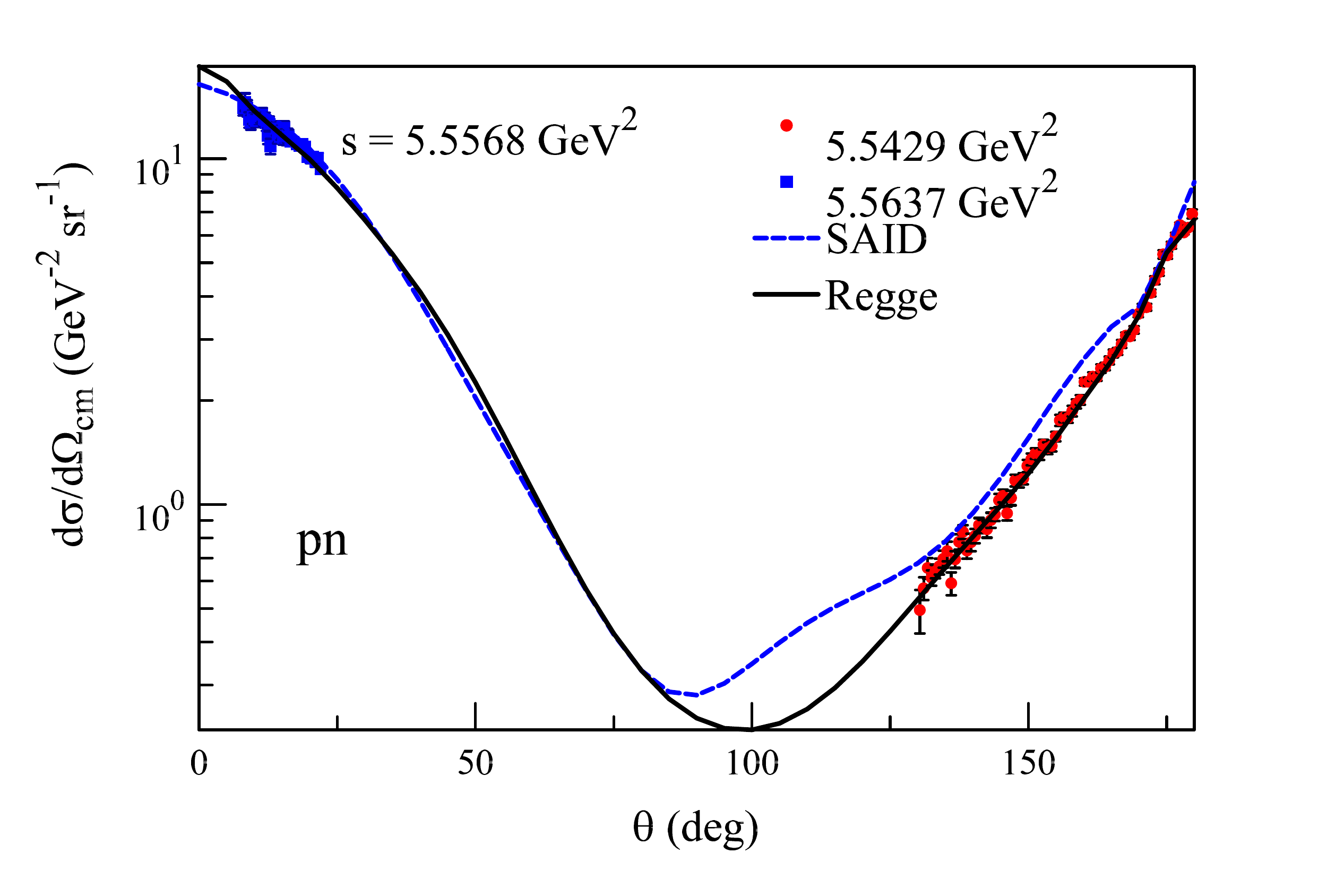}}
\caption{(color online) Center of momentum $pn$ elastic cross sections calculated using both SAID and Regge methods. Data close to the chosen value of $s$ are also displayed.
}
\label{fig:dsigdOmega}
\end{figure}

For the E1210003 kinematics the square of the invariant mass of the final state is $s=5.5568\ {\rm Gev^2}$ which is well above the pion-production threshold and beyond the range where meson exhange models have been capable of reproducing the $NN$ cross sections and spin observables. This means that it is only possible to describe the final state interactions in terms of fits of parameterized amplitudes that are fit to available $NN$ scattering data. Two methods are available that contain the full spin dependence of the amplitudes. The first of these is the use of the helicity amplitudes that are available from SAID. For the $pn$ amplitudes these are reliable only up to about $T_{lab}=1.3\ {\rm GeV}$ or $s=5.9675\ {\rm GeV^2}$. The second method is a fit to $NN$ cross sections and spin observables from $s=5.4\ {\rm GeV^2}$ to $s=4000\ {\rm GeV^2}$ using a Regge model parameterization. The E1210003 kinematics are therefore in a region where both methods may be used. Figure \ref{fig:dsigdOmega} shows the c.m. differential $pn$ elastic cross sections at the value of $s$ for the E1210003 kinematics.
Some care should be taken in judging the relative quality of the two methods based on this single figure. In fitting the differential cross sections the normalizations are generally allowed to float due to the difficulty of experimentally determining absolute normalization. The data shown have been rescaled as required by the Regge model fit. A careful comparison would also include comparisons of spin observables. At this point it is reasonable to assume that either method produces results that can be reasonably used in calculating the deuteron electrodisintegration cross sections.

\section{Calculations}\label{sec:Calculations}

We now have 8 possible choices of wave function, 3 choices of electromagnetic form factors and 2 choices for the final state interaction as summarized in Tab. \ref{tab:model_inputs}. This means that there are 24 possible calculations for the PWIA given by diagram Fig. \ref{fig:IA_diagrams}(a) and 48 possible calculations for the complete IA given by all of the diagrams in Fig. \ref{fig:IA_diagrams}.

\begin{figure}
{\includegraphics[height=3in]{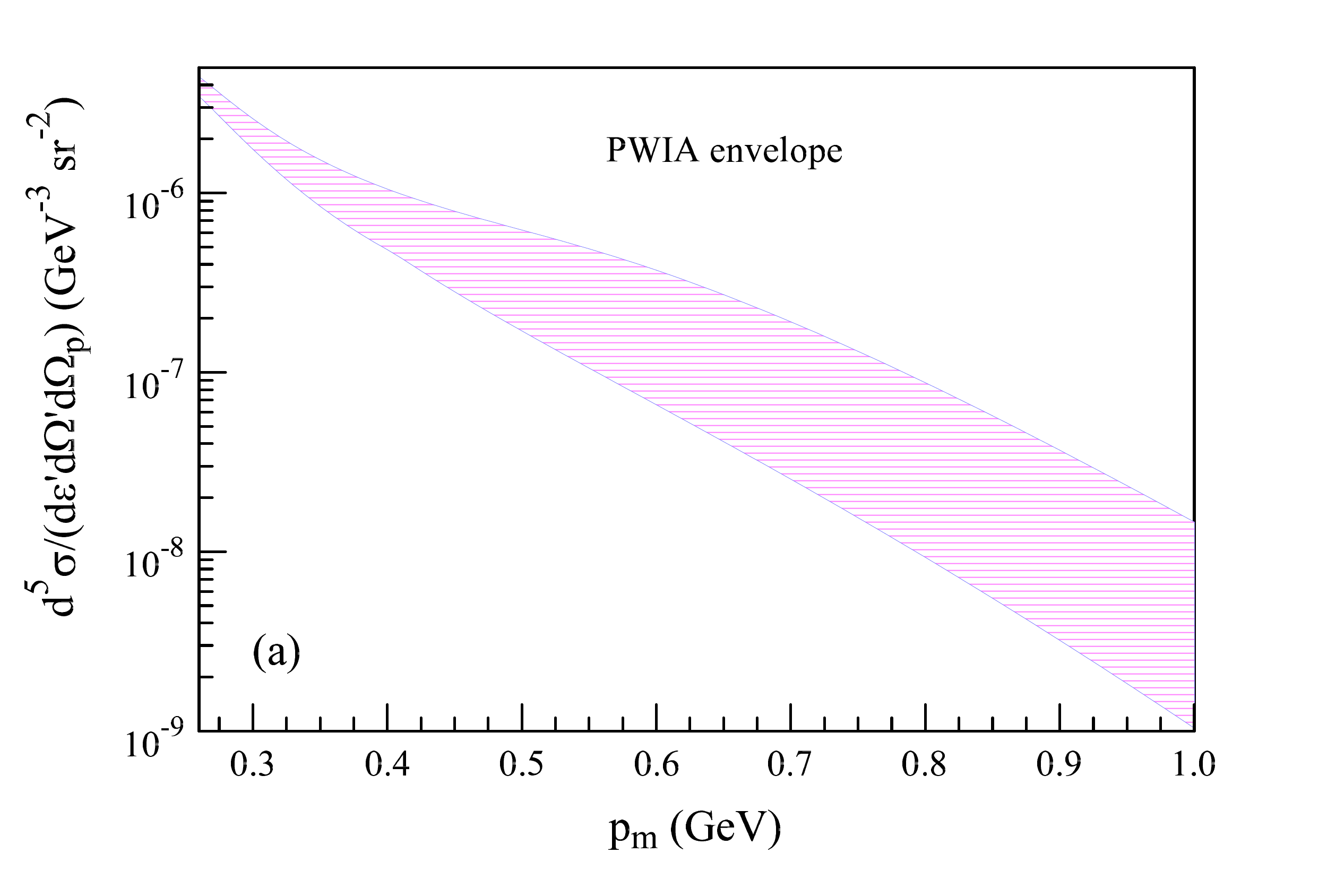}}
{\includegraphics[height=3in]{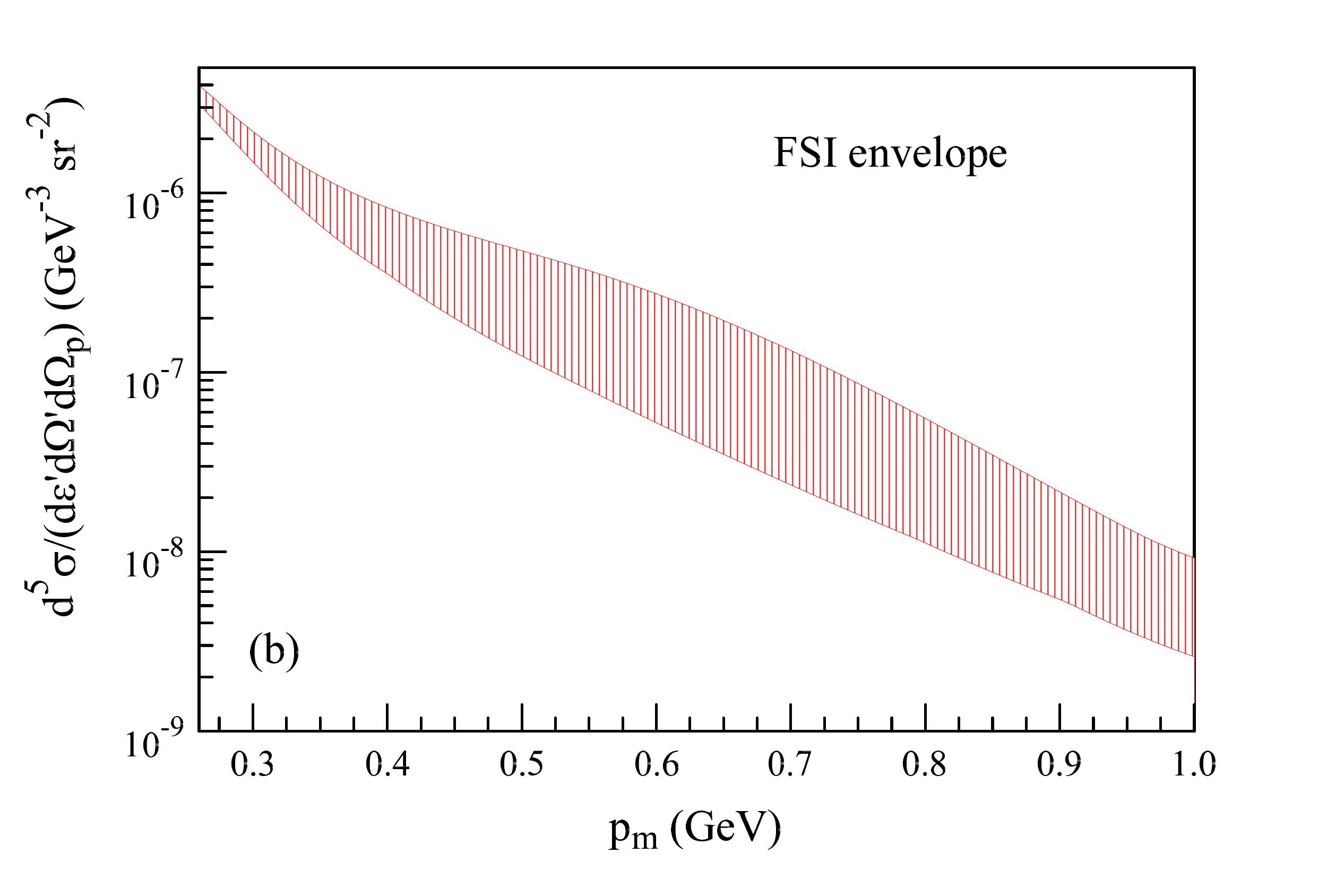}}
\caption{(color online) The envelope containing all 24 calculations of the cross section in PWIA is shown in panel (a). The corresponding envelope containing all 48 of the complete calculations represented by the diagrams of Fig. \ref{fig:IA_diagrams} is shown in panel (b).
}
\label{fig:pwia_FSI_envelope}
\end{figure}

Since the number of calculations in each case is large we choose to plot the envelopes containing all 24 of the PWIA calculations and all 48 of the full IA calculations for the E1210003 kinematics in Fig. \ref{fig:pwia_FSI_envelope}. That is in each case for each $p_m$ we determine the largest and smallest values given by the set of calculations giving the boundaries of the shaded areas or envelopes. Note that the envelope for the PWIA calculatons, Fig. \ref{fig:pwia_FSI_envelope}(a) increases in width with increasing $p_m$ and covers a range of more than an order of magnitude at $p_m=1\ {\rm GeV}$ which is in agreement with the range of momentum distributions shown in Fig. \ref{fig:mom_dist}. We would like to point out that the inclusion of FSIs reduces the width of the envelope at high missing momenta as the FSIs redistribute strength from the low-momentum part of the wave function to high momenta, and the PWIA envelope 
is narrower at low missing momentum.

Figure \ref{fig:pwia_FSI_envelope}(b) shows the envelope containing the 48 cross section calculations with FSI. Note that above approximately $p_m=0.65\ {\rm GeV}$ the envelope for the FSI calculations begins to narrow and covers a significantly smaller range at $p_m=1\ {\rm GeV}$. 

\subsection{Choice of Kinematics}

\begin{figure}
{\includegraphics[height=3in]{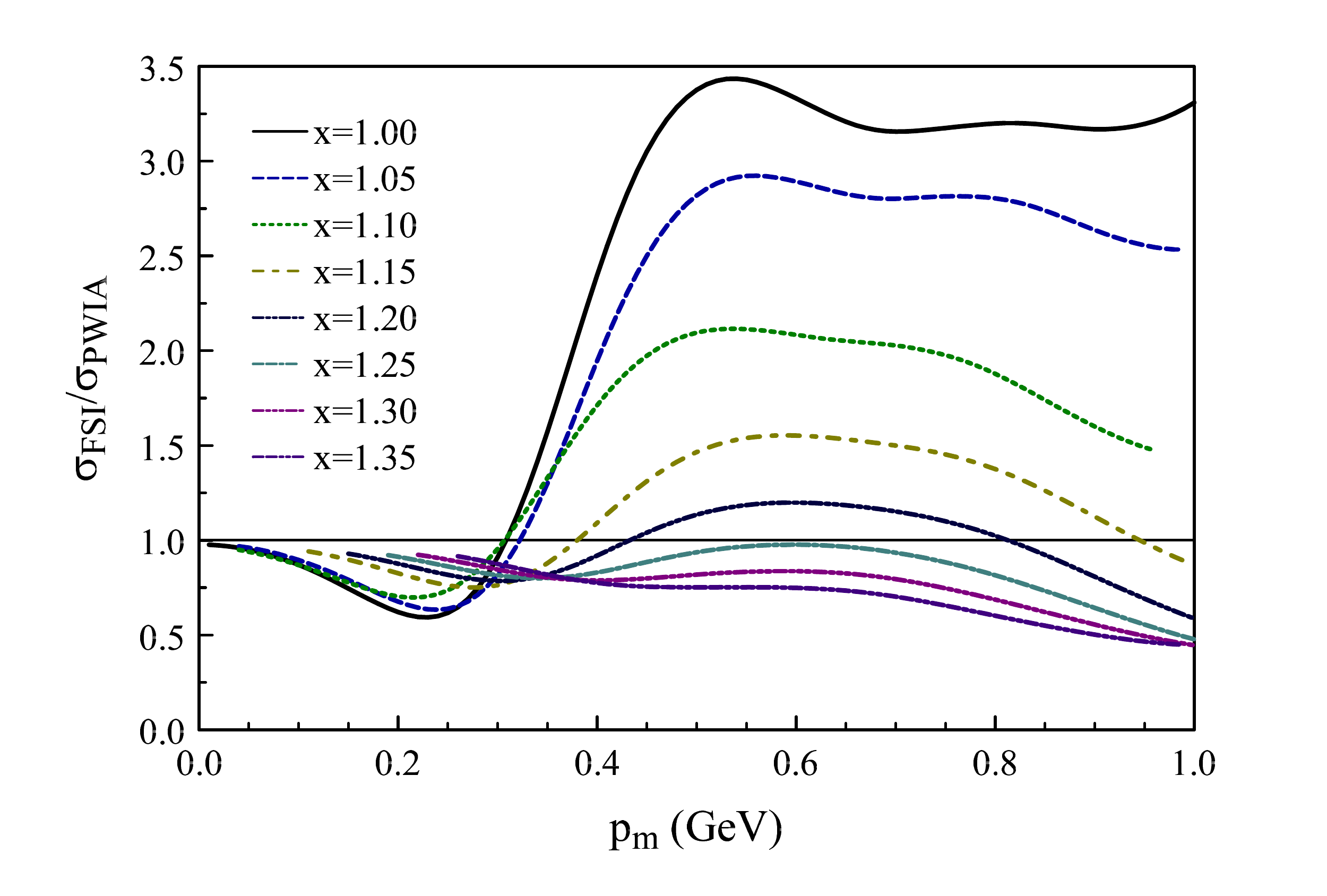}}
\caption{(color online) The ratio of FSI to PWIA cross sections for wave function IIB, the GKex05 electromagnetic form factors at $\varepsilon=11\ {\rm GeV}$, $Q^2=4.25\ {\rm GeV^2}$ and a range of values for $x$.
}
\label{fig:ratio_x}
\end{figure}

\begin{figure}
{\includegraphics[height=3in]{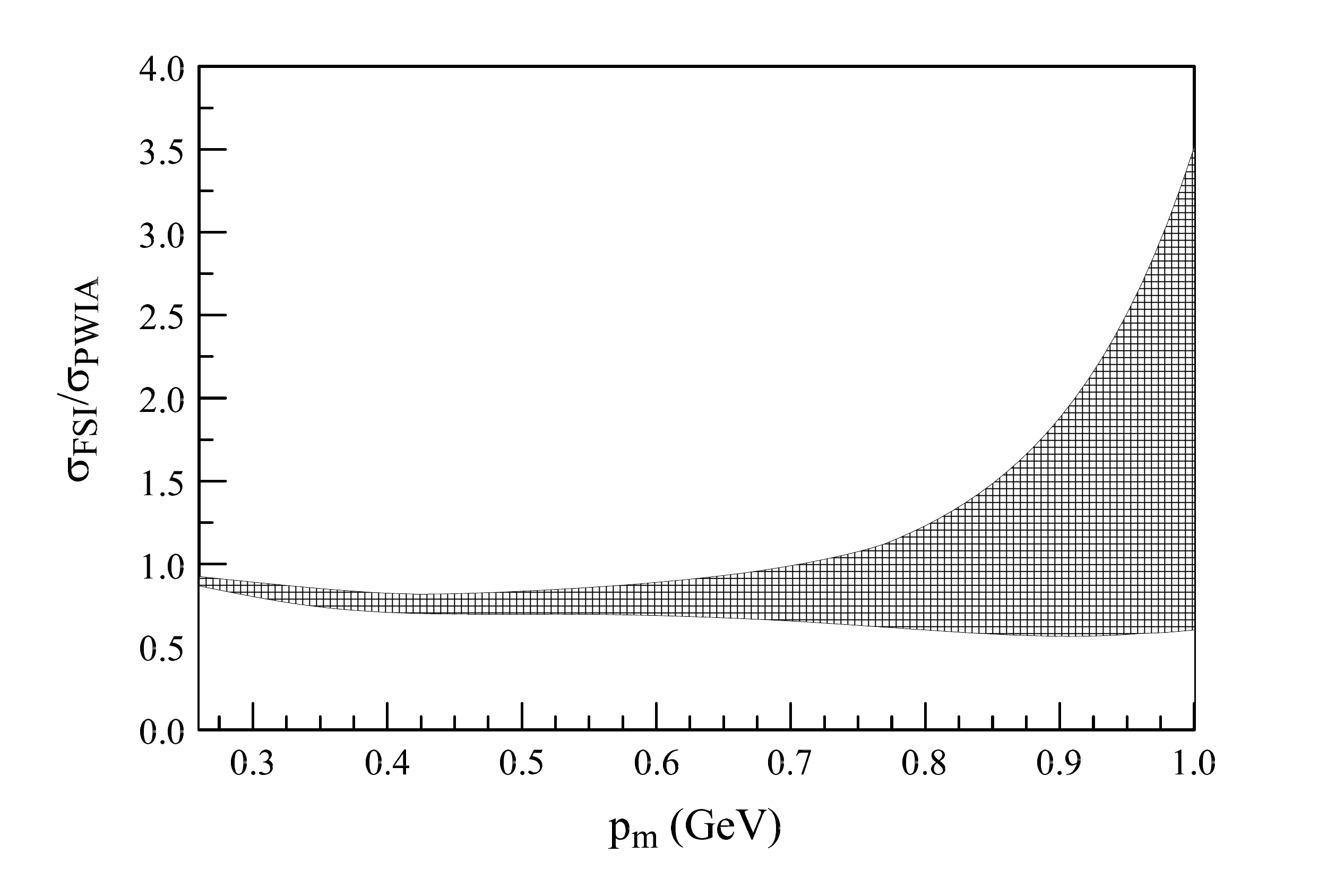}}
\caption{(color online) The envelope containing the ratio FSI to PWIA cross sections for all 48 calculations is shown by the shaded band.
}
\label{fig:ratio_envelope}
\end{figure}

To argue that the reduced cross section is a rough representation of the deuteron momentum distribution requires that a region of kinematics must be found where the role of FSI is minimal. This is the approach used in \cite{Boeglin_highmom} and for the E1210003 kinematics. The ability to do this using the IIB wave functions, the GKex05 electromagnetic form factors and the Regge model FSI is shown in Fig. \ref{fig:ratio_x}. Here 
we show the ratio of the cross section for the full IA to the corresponding PWIA. In this figure the incident electron energy is $\varepsilon=11\ {\rm GeV}^2$, $Q^2=4.25\ {\rm GeV^2}$ and $x$ is allowed to vary from 1 to 1.35 in steps of 0.05. A ratio of 1 would indicate that the FSI had no effect at a given kinematics. 
For all values of $x$ this ratio is below 1 for $p_m\lesssim 0.3\ {\rm GeV}$. The ratios then increase to above 1 for $1\leq x\leq 1.2$ with the magnitude decreasing as $x$ increases. For $x\geq 1.25$ the ratio remains below 1. In this case, it would seem that the choice of $x=1.25$ would tend to minimize the role of FSI while the E1210003 kinematics, at $x = 1.35$, would increase the role of FSI. Since the choice of optimal kinematics relies on calculation, it is not surprising that the choice is model dependent.
  The extent of this problem can be shown by plotting the envelope containing the ratio FSI to PWIA cross sections for all 48 cases at the E1210003 kinematics. This envelope is shown by the shaded area in Fig. \ref{fig:ratio_envelope}.
The large upper value of the ratio at $p_m=1\ {\rm GeV}$ is the result of calculations using the CD Bonn wave functions which produce the lower values of the PWIA and FSI calculations at large $p_m$. Since the final state interactions tend to raise the cross section in this region and the PWIA cross sections for CD Bonn are small, the ratio of cross sections then becomes large.

\begin{figure}
{\includegraphics[height=3in]{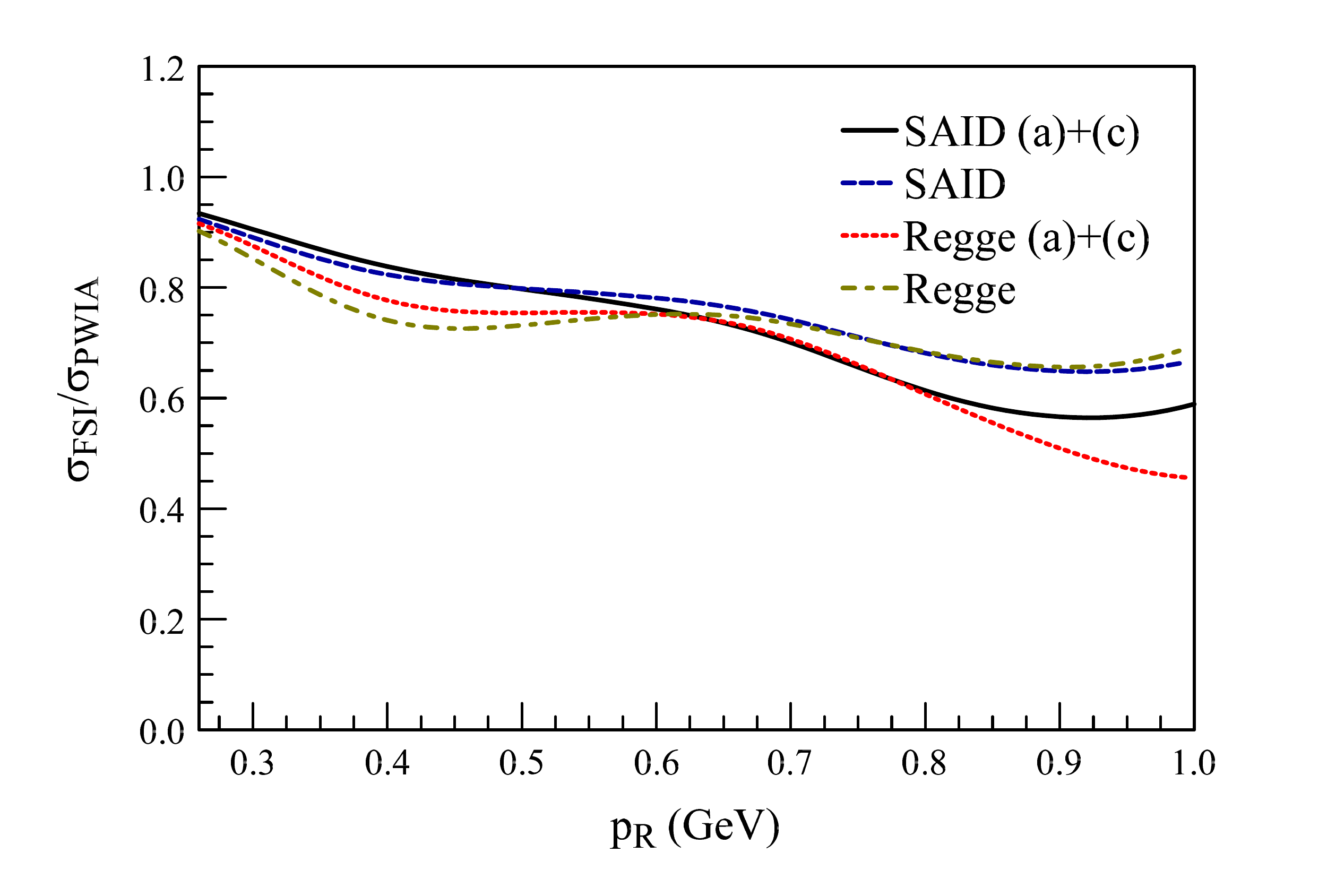}}
\caption{(color online) Ratio of FSI to PWIA cross sections for wave function IIB and the AMT electromagnetic form factors at the E1210003 kinematics for both SAID and Regge FSI. Curves labeled with (a)+(c) contain only the proton contributions to the IA (diagrams (a)and (c) of Fig. \ref{fig:IA_diagrams}, while those without the label also contain contributions of from the neutron given by diagrams (b) and (d) of Fig. \ref{fig:IA_diagrams}.
}
\label{fig:ratio_AMT}
\end{figure}

The cause of the narrowing of the range of FSI cross sections at large $p_m$ is illustrated by Fig. \ref{fig:ratio_AMT}. In this figure the ratio of FSI to PWIA cross sections for the IIB wave function and the AMT electromagnetic form factors is shown for the E1210003 kinematics. The curves labeled SAID (a)+(c) or Regge (a)+(c)  have contributions only from diagrams (a) and (c) of Fig. \ref{fig:IA_diagrams} where the electron scatters from the proton only. The curves labeled SAID or Regge have contributions from all of the diagrams in Fig. \ref{fig:IA_diagrams} including contributions where the electron scatters from both the proton and the neutron. While the inclusion of the neutron contributions is relatively small at lower $p_m$, at larger $p_m$ they have the effect of causing the ratios for the SAID and Regge FSI to become very close in value. The neutron contributions are then responsible for narrowing the range of the FSI calculations at large $p_m$. This indicates that the complete IA must be used in calculation of cross sections at the E1210003 kinematics.

The model dependence of the choice of optimal kinematics along with the substantial range in the values of the cross sections at large $p_m$ implies that a method for obtaining momentum distributions from data be found that is less sensitive to the choice of optimal kinematics and includes information about the range of possible calculations.  We will describe one possible approach to this problem in the following section.

\section{A New Method for Extracting the Deuteron Momentum Distribution}\label{sec:NewMethod}

In formulating a new approach to obtaining the deuteron momentum distribution it should include information about how well the reduced cross section for each represents the actual momentum distribution calculated directly from the wave functions used in the model. It should also include  an estimate of the theory error associated with the wide range of possible calculations that can be produced by the acceptable range of wave functions, electromagnetic form factors and final state interactions that can be combined to produce the calculations. It should then take into account the fact that the momentum distribution is not an observable quantity. To accomplish this we propose the following procedure.

Our goal is to provide a procedure for the extraction of an experimental momentum distribution,
$n_{exp}(p)$. It can be obtained in the following way:

\begin{equation}
n_{exp}(p)=\frac{\sigma_{exp}(p)}{k \sigma_{ep}(p)}-\left<\xi_{th}(p)\right>\pm \delta\xi_{th}(p)\pm\delta\sigma_{red}(p)\,.
\end{equation}

Here, $\sigma_{exp}(p)$ is the experimentally measured cross section, and $k \sigma_{ep}$ is the 
factor that is used to extract the reduced cross section, see our description of the method 
in Appendix \ref{app:factor}. The reduction factor contains the (off-shell) electron-proton
cross section $\sigma_{ep}$, which requires an electromagnetic form factor. This form factor is 
chosen from one of the available parameterizations. The term $\delta\sigma_{red}(p)$ is the 
experimental error.

The other two terms account for the theoretical difference between the calculated reduced cross section
and the corresponding calculated momentum distribution, and its theoretical error, $\delta\xi_{th}(p)$.
These two quantities are obtained as follows: 
for each of the $N =48$ possible calculations, labeled $i$, we calculate the theoretical quantity
\begin{equation}
\xi_{th_i}(p)=\frac{\sigma_{eD_i}(p)}{k\sigma_{ep}(p)}-n_{th_i}(p)
\end{equation}
for a range of values of $p$. The first term is the calculated reduced cross section which we calculate here using the method presented in Appendix \ref{app:factor} with the same electromagnetic form factors used in $\sigma_{ep}$ for the extraction of the experimental reduced cross section, for all 48 variations
of the theoretical calculation.
%
This quantity therefore represents the difference between the reduced cross section and the actual momentum distribution for the wave functions used in the calculation. The average value of this difference for all calculations can be calculated as
\begin{equation}
\left<\xi_{th}(p)\right>=\frac{1}{N}\sum_{i=1}^N\xi_{th_i}(p)\label{eq:xith}
\end{equation}
and the average of the square of the difference as
\begin{equation}
\left<\xi^2_{th}(p)\right>=\frac{1}{N}\sum_{i=1}^N\xi^2_{th_i}(p)\,.
\end{equation}
The standard deviation of this difference is
\begin{equation}
\delta\xi_{th}(p)=\sqrt{\left<\xi_{th}^2(p)\right>-\left<\xi_{th}(p)\right>^2}
\end{equation}
and can be taken as an approximate measure of the theory error.

%

\section{Results}\label{sec:Results}

Preliminary examples of how this method may work can be obtained by using selected cross sections from the 48 used in this work as pseudo-data and determining how well the procedure reproduces the corresponding theoretical momentum distributions.

\begin{figure}
{\includegraphics[height=3in]{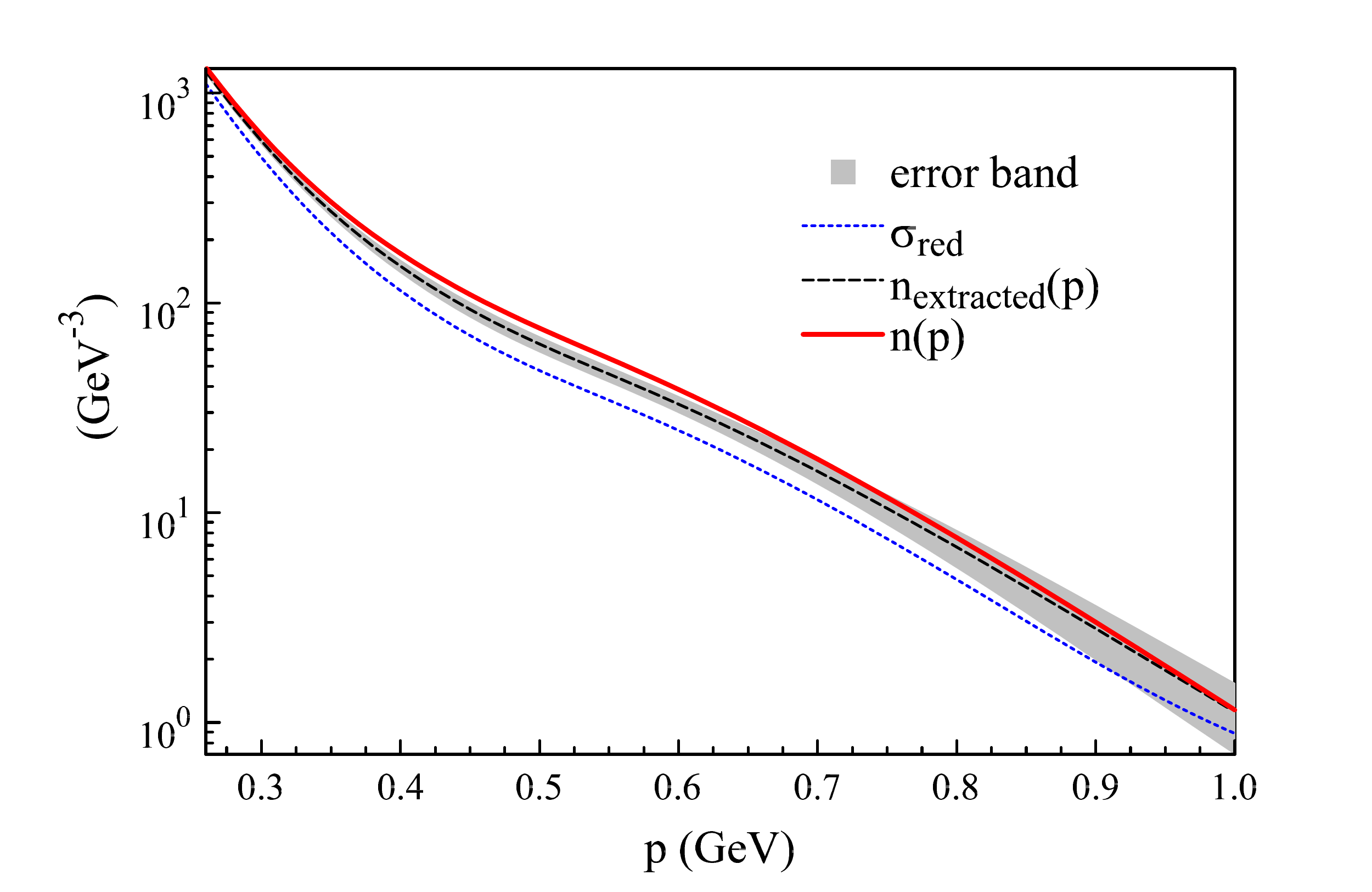}}
\caption{(color online) This figure uses the calculation of the cross section with the WJC 2 wave functions, the GKex05 electromagnetic form factor and the Regge FSI as pseudo-data. The dotted line is the reduced cross section using the AMT form factor in $\sigma_{ep}$, the dashed line is the extracted momentum distribution using the procedure described above with a shaded band representing the theoretical error. The solid line is the momentum distribution for the WJC 2 wave functions.
}
\label{fig:wjc2}
\end{figure}

In Fig. \ref{fig:wjc2} the pseudo-data are represented by the calculated cross section for the WJC 2 wave function, the GKex05 electromagnetic form factors and the Regge model FSI. The reduced cross section is calculated using the factorization procedure of Appendex \ref{app:factor} with the AMT proton form factors used in the reduction factor $\sigma_{ep}$. The reduced cross section is represented by the dotted line and the central value of the extracted momentum distribution using the procedure of Section \ref{sec:NewMethod} is represented by the dashed line and the theoretical error is represented by the shaded band. The calculated momentum distribution for the WJC 2 wave functions is given by the solid line. 
At momenta above 0.7 GeV the extracted momentum distribution and the calculated distribution agree within the theoretical error.

\begin{figure}
{\includegraphics[height=3in]{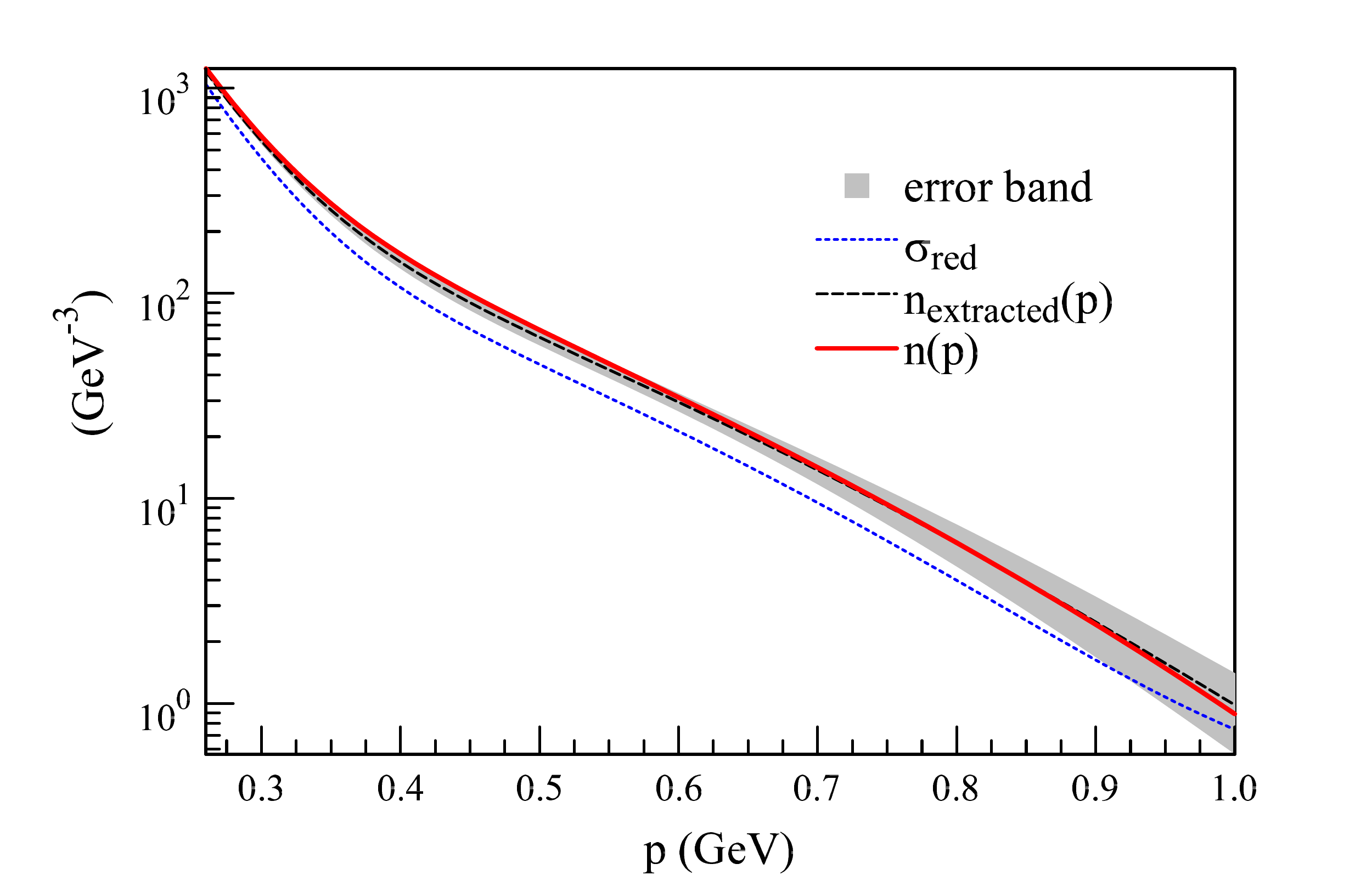}}
\caption{(color online) Same as Fig. \ref{fig:wjc2} but using the AV18 wave function.
}
\label{fig:av18}
\end{figure}

Figure \ref{fig:av18} uses the calculation for the AV18 wave function, with the same electromagnetic form factors and FSI as the previous figure, as pseudo-data. In this case the extracted and calculated momentum distributions are in excellent agreement for large momenta and are well within the theoretical error.

\begin{figure}
{\includegraphics[height=3in]{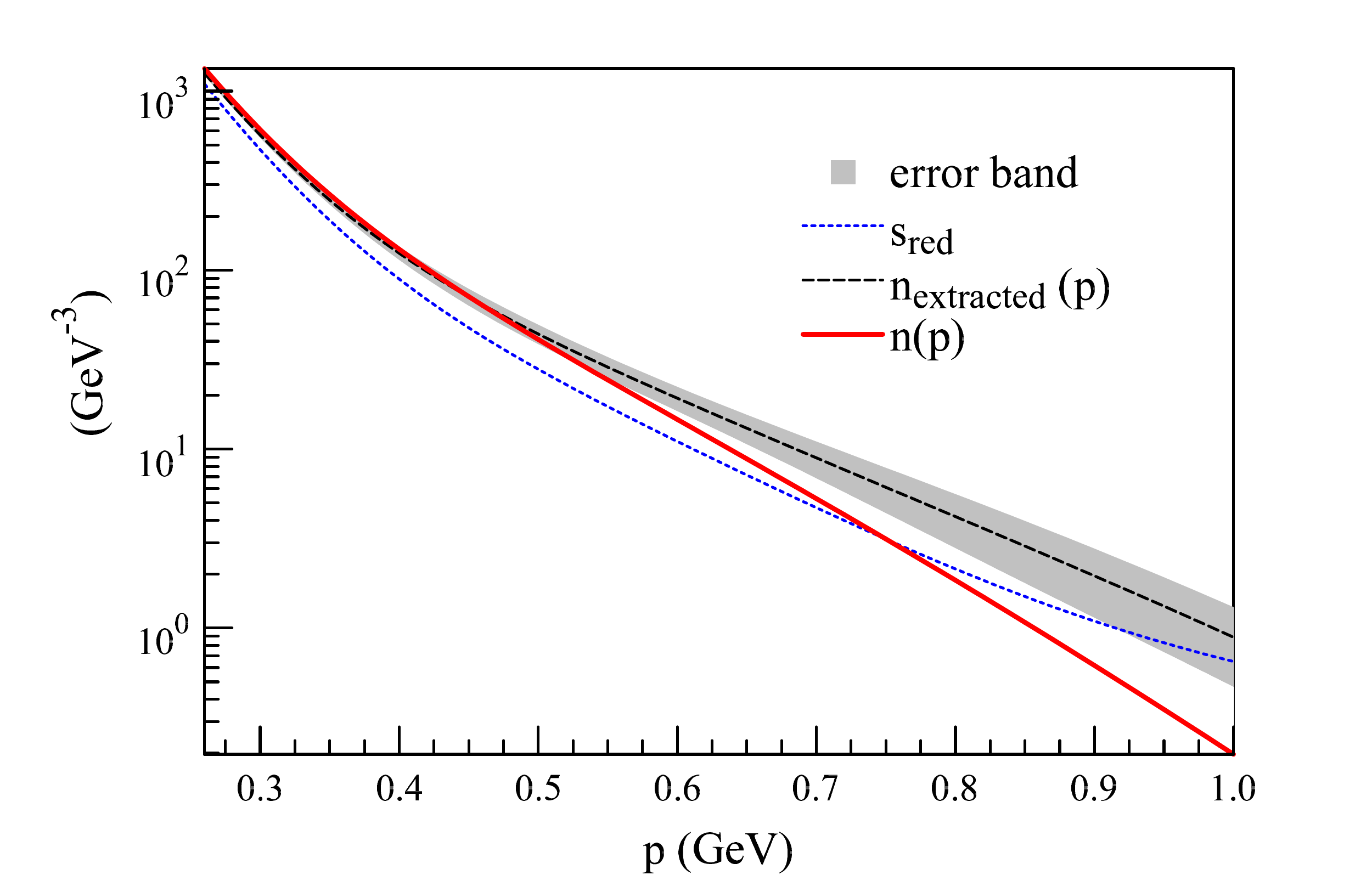}}
\caption{(color online) Same as Fig. \ref{fig:wjc2} but using the CD Bonn wave function.
}
\label{fig:cdb}
\end{figure}

\begin{figure}
{\includegraphics[height=3in]{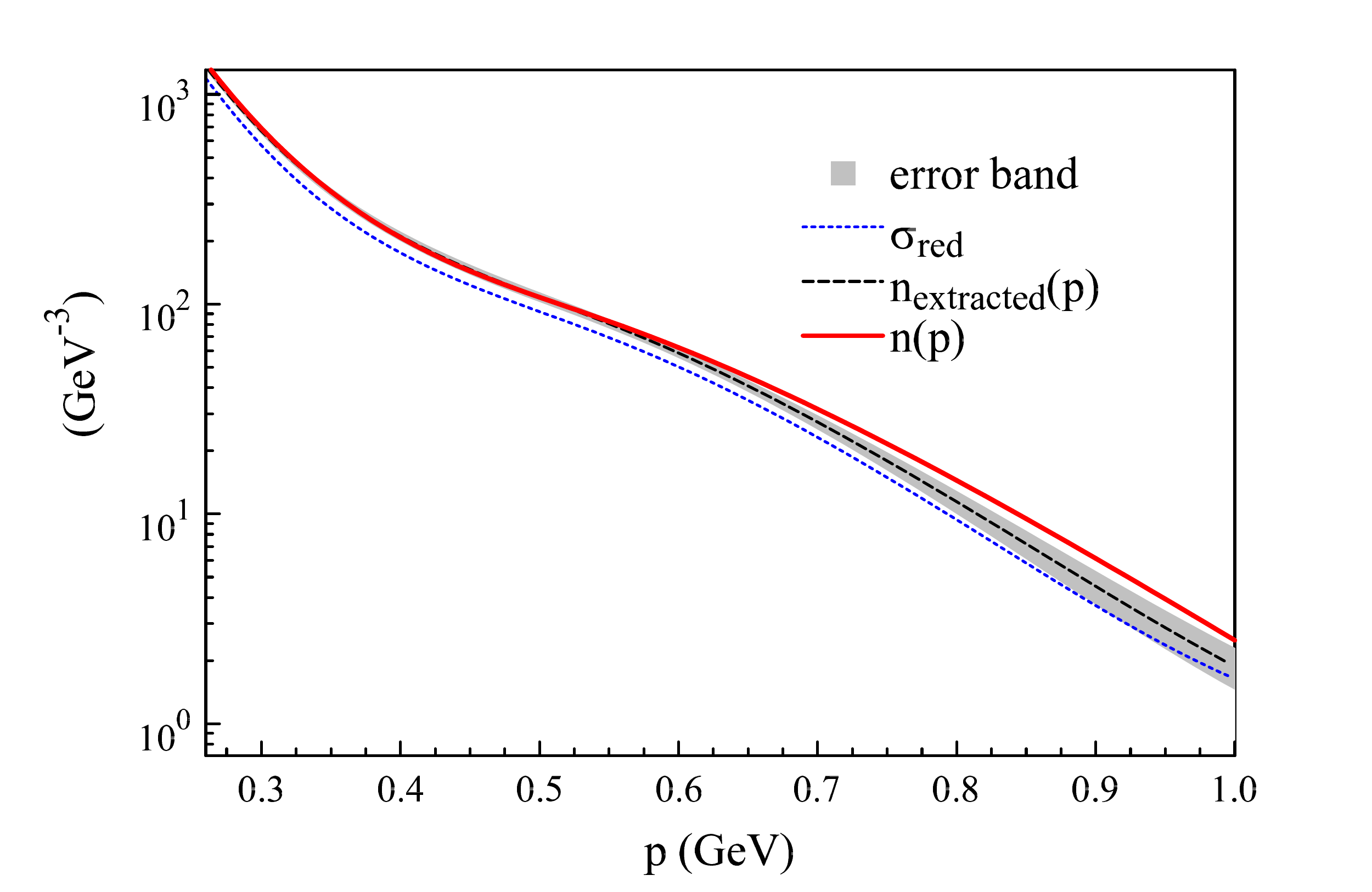}}
\caption{(color online) Same as Fig. \ref{fig:wjc2} but using the WJC 1 wave function.
}
\label{fig:wjc1}
\end{figure}

The calculations are repeated using the CD Bonn wave functions in Fig. \ref{fig:cdb}. In this case the extracted momentum distribution is much larger than the calculated distribution. The CD Bonn potential is by far the softest of those used here. The integration over the FSI moves strength from lower momentum to higher momentum which causes a much larger effect for the softer wave functions. As a result, the reduced cross section is much larger at large momentum indicating that the ratio of the full calculation to the PWIA is much larger than 1. This means that the approach presented here will tend to give an upper bound on the momentum distribution for all but the softest of potentials. 
As a contrast to the case of the CD Bonn wave functions, Fig. \ref{fig:wjc1} uses the WJC 1 wave function which is the hardest of those used in this work. In this case the calculated momentum distribution is larger than the extracted momentum distribution, but is well within the range implied by statistics. In contrast to the previous case the effect of the FSI on the extracted momentum distribution is much smaller than is the case for the CD Bonn wave functions.

\section{Summary and Conclusions}\label{sec:Summary}

The goal of this paper is to find a measure of the theoretical uncertainties in the extraction of momentum distributions
from experimental data that are due to model {\sl inputs}. Model inputs --- electric and magnetic form factors, wave functions, and nucleon-nucleon scattering amplitudes --- are necessary for all theoretical calculations. There are several versions of these available in the literature, and all of them are widely used. So, completely apart from
the theoretical model used to describe the reaction mechanism of the $^2H(e,e'p)$ reaction, there will be uncertainties
involved that stem from these inputs. 

We have mimicked the experimental data with a set of calculations, and then used all 48 possible input combinations
to extract the momentum distribution, leading to an error band. We performed our calculations at the kinematics for the
planned Jefferson Lab E1210003 experiment. In all studied cases, the error band has a reasonable width that tends to increase with higher missing momentum. The increase in uncertainty at higher momentum can in part be attributed to the
contribution of graphs with final state interactions and contributions from virtual photon absorption on the neutron. 

In most of our examples, the band that represents the theory input error around the extracted momentum distribution
includes the momentum distribution $n_{th}$ consistent with the calculation used to generate the pseudo-data in the first place. The approach presented here will tend to give an upper bound on the momentum distribution for all but the softest of potentials. 
We are confident that the method for a calculation of the theoretical error band provided in this paper will be
very helpful for the analysis of the forthcoming high-precision data from Jefferson Lab's 12 GeV upgrade.

{\bf Acknowledgments}:  This material is based upon work supported by the U.S. Department of Energy, Office of Science, Office of Nuclear Physics under contract DE-AC05-06OR23177, and was
supported in part by funds provided by the U.S. Department of Energy
(DOE) under cooperative research agreement under No.
DE-AC05-84ER40150. One of the authors (SJ) was supported in part by NSF grant PHY-1306250.
The authors would like to thank Rocco Schiavilla and Franz Gross for providing numerical values of the wave functions for the various potentials.

\appendix

\section{Cross section factorization and the momentum distribution}\label{app:factor}

\begin{figure}
\centerline{\includegraphics[height=2in]{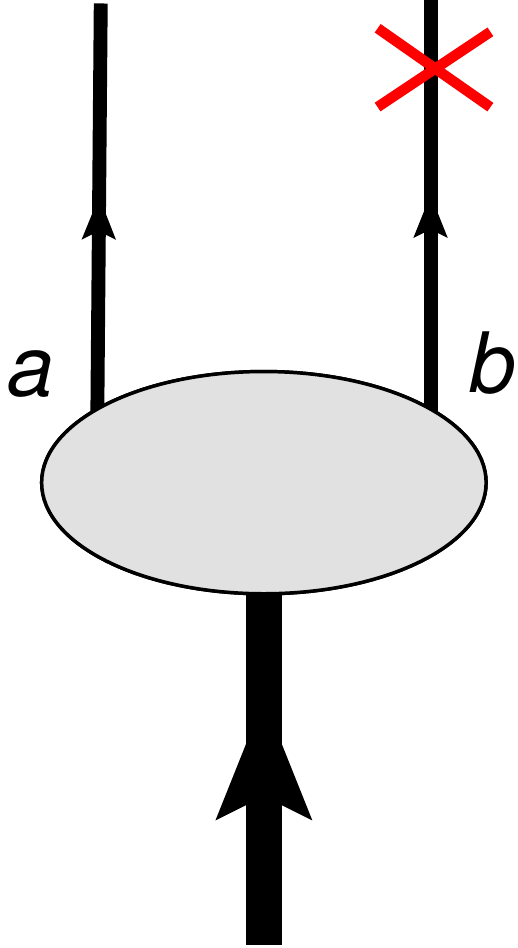}}
\caption{(color online) Diagram representing the half-onshell deuteron vertex function.
}\label{fig:vertex}
\end{figure}

The extraction of the deuteron momentum distribution from measured cross sections depends upon the assumption that the cross section can be factored into a factor due to scattering on an off-shell proton and a factor equal to the momentum density distribution. This factorization becomes more complicated when FSI are introduced and becomes somewhat dependent upon the theoretical formalism used. The factorization procedure used here can be obtained directly from consideration of the PWIA contribution described by Fig. \ref{fig:IA_diagrams}a. For this diagram both of the final state nucleons are on-shell which implies that the vertex function has particle 2 on shell as well. This is represented by Fig. \ref{fig:vertex}. 
For either the full Bethe-Salpeter equation or the spectator, or Gross equation, the half-off-shell vertex function can be written as 
\begin{eqnarray}
\Gamma_{\lambda_d}(p_2,P)&=&\left[ g_1(p_2^2,p_2\cdot
P)\gamma\cdot\xi_{\lambda_d}(P) +g_2(p_2^2,p_2\cdot P)
\frac{p\cdot\xi_{\lambda_d}(P)}{m}\right.\nonumber\\
&&\left.-\left(g_3(p_2^2,p_2\cdot P)\gamma\cdot\xi_{\lambda_d}(P)
+g_4(p_2^2,p_2\cdot
P)\frac{p\cdot\xi_{\lambda_d}(P)}{m}\right)\frac{\gamma\cdot
p_1+m}{m}\right]C\,.\label{eq:vertex_function}
\end{eqnarray}
where $\xi_{\lambda_d}(P)$ is the deuteron polarization four-vector, $C$ is the charge conjugation matrix and the invariant functions $g_i(p_2^2,p_2\cdot P)$ are given by
\begin{eqnarray}
g_1(p_2^2,p_2\cdot P)&=&\frac{2E_{p_R}-M_d}{\sqrt{8\pi}}\left[ u(k)-\frac{1}{\sqrt{2}}w(k)+\sqrt{\frac{3}{2}}\frac{m}{{p_R}}v_t({p_R})\right]\nonumber\\
g_2(p_2^2,p_2\cdot P)&=&\frac{2E_{p_R}-M_d}{\sqrt{8\pi}}\left[ \frac{m}{E_{p_R}+m}u({p_R})+\frac{m(2E_{p_R}+m)}{\sqrt{2}{p_R}^2}w({p_R})+\sqrt{\frac{3}{2}}\frac{m}{{p_R}}v_t({p_R})\right]\nonumber\\
g_3(p_2^2,p_2\cdot P)&=&\sqrt{\frac{3}{16\pi}}\frac{m E_{p_R}}{{p_R}}v_t({p_R})\nonumber\\
g_4(p_2^2,p_2\cdot P)&=&-\frac{m^2}{\sqrt{8\pi}M_d}\left[(2 E_{p_R}-M_d)\left(\frac{1}{E_{p_R}+m}u({p_R}) -\frac{E_{p_R}+2m}{\sqrt{2}{p_R}^2}w({p_R})\right)+\frac{\sqrt{3}M_d}{{p_R}}v_s({p_R})\right]\,,\nonumber\\\label{eqn:g1to4}
\end{eqnarray}
Here, the scalar ${p_R}$ is defined as
\begin{equation}
{p_R}=\sqrt{\frac{(P\cdot p_2)^2}{P^2}-p_2^2}
\end{equation}
and is the magnitude of the neutron three-momentum in the deuteron rest
frame and the corresponding energy is
\begin{equation}
E_{p_R}=\sqrt{{p_R}^2+m^2}\,.
\end{equation}
The functions $u({p_R})$, $w({p_R})$, $v_s({p_R})$ and $v_t({p_R})$ are the s-wave,
d-wave, singlet and triple p-wave radial wave functions of
the deuteron in momentum space.  

For convenience, the half-off-shell deuteron wave function can be defined
as
\begin{equation}
\psi_{\lambda_d,s_2}(p_2,P)=G_0(P-p_2)
\Gamma^T_{\lambda_d}(p_2,P)\bar{u}^T(\bm{p}_2,s_2)\,.
\end{equation}
We choose to normalize this wave function such that in the deuteron
rest frame
\begin{equation}
\sum_{s_2}\int\frac{d^3p_2}{(2\pi)^3}\frac{m}{E_{p_2}}\bar{\psi}_{\lambda_d,s_2}(p_2,P)
\gamma^0 \psi_{\lambda_d,s_2}(p_2,P)=1\,,
\end{equation}
which is correct only in the absence of energy-dependent kernels.
This results in the normalization of the radial wave functions given by
\begin{equation}
\int_0^\infty
\frac{dpp^2}{(2\pi)^3}\left[u^2(p)+w^2(p)+v_t^2(p)+v_s^2(p)\right]=1\,.
\end{equation}

The plane wave contribution to the current matrix element represented by Fig. \ref{fig:IA_diagrams}a can then be written as
\begin{equation}
\left<\bm{p}_1s_1;\bm{p}_2s_2\right|J^\mu_{(1)}\left|\bm{P}\lambda_d\right>_a=
-\bar{u}(\bm{p}_1,s_1)\Gamma^\mu(q)\psi_{\lambda_d,s_2}(p_2,P)\,,
\end{equation}
where the one-body nucleon electromagnetic current operator is chosen to be of the Dirac-plus-Pauli form
\begin{equation}
\Gamma^\mu(q)=F_1(Q^2)\gamma^\mu+\frac{F_2(Q^2)}{2m}i\sigma^{\mu\nu}q_\nu\,.
\end{equation}

The PWIA response tensor is then
\begin{align}
W^{\mu\nu}_{aa}=&\frac{1}{3}\sum_{s_1,s_2,\lambda_d}\overline{\psi}_{\lambda_d,s_2}(p_2,P)\Gamma^\mu(-q)u(\bm{p}_1,s_1)\bar{u}(\bm{p}_1,s_1)\Gamma^\nu(q)\psi_{\lambda_d,s_2}(p_2,P)\nonumber\\
=&{\rm Tr}[\Gamma^\mu(-q)\Lambda_+(\bm{p}_1)\Gamma^\nu(q)N(p_2,P)]\,,
\end{align}
where the momentum distribution operator is given by
\begin{equation}
N(p_2,P)=\frac{1}{16\pi}\left[\frac{P\cdot p_2}{M_d^2m_N}\gamma\cdot P n_{tv}({p_R})-\left(\frac{\gamma\cdot p_2}{m_n}-\frac{P\cdot p_2}{M_d^2m_N}\gamma\cdot P\right)n_{sv}({p_R})+n_s({p_R})\right]
\end{equation}
with three scalar momentum distributions defined as
\begin{align}
n_{tv}({p_R})=&u^2({p_R})+w^2({p_R})+v_t^2({p_R})+v_s^2({p_R})\\
n_{sv}({p_R})=&u^2({p_R})+w^2({p_R})-v_t^2({p_R})-v_s^2({p_R})\nonumber\\
&-\frac{2m_N}{\sqrt{3}{p_R}}((u({p_R})+\sqrt{2}w({p_R}))v_s({p_R})-(\sqrt{2}u({p_R})-w({p_R}))v_t({p_R}))\\
n_s({p_R})=&u^2({p_R})+w^2({p_R})-v_t^2({p_R})-v_s^2({p_R})\nonumber\\
&+\frac{2 {p_R}}{\sqrt{3}m_N}((u({p_R})+\sqrt{2}w({p_R}))v_s({p_R})-(\sqrt{2}u({p_R})-w({p_R}))v_t({p_R}))\,.
\end{align}
Note that only the time-like-vector momentum distribution $n_{tv}$ is related to the normalization condition such that
\begin{equation}
\int_0^\infty
\frac{dpp^2}{(2\pi)^3}n_{tv}(p)=1\,.\label{eq:normalize}
\end{equation}

In the absence of relativistic p-wave contributions all three momentum distributions are the same and we can define
\begin{equation}
n_+({p_R})=n_{tv}({p_R})=n_{sv}=n_s({p_R})=u^2({p_R})+w^2({p_R})\,.
\end{equation}
which is the usual nonrelativistic momentum distribution. In this case the momentum density operator becomes
\begin{equation}
N_+(p_2,P)=\frac{1}{8\pi}\Lambda_+(\bm{p})n_+(p)\,,
\end{equation}
where
\begin{equation}
p=(\sqrt{p^2+m_N^2},\bm{p})\,.
\end{equation}

The PWIA response tensor then becomes
\begin{equation}
W^{\mu\nu}_{aa}=\frac{1}{8\pi}{\rm Tr}[\Gamma^\mu(-q)\Lambda_+(\bm{p}_1)\Gamma^\nu(q)\Lambda_+(\bm{p})]n_+(p)\,,\label{eq:W_factored}
\end{equation}
which clearly factors into a contribution composed of an off shell single-nucleon contribution and the positive-energy momentum distribution.

The off-shell four momentum of the struck proton is given in the rest frame, by
\begin{equation}
k=P-p_2=(M_d,\bm{0})-(E_p,-\bm{p})=(M_d-E_p,\bm{p})=p+(M_d-2 E_p,\bm{0})=p+\Delta\,,
\end{equation}
where
\begin{equation}
\Delta=(M_d-2E_p,\bm{0})=(\delta,\bm{0})\,.
\end{equation}
Four-momentum conservation requires that 
\begin{equation}
k+q=p+\Delta+q=p_1
\end{equation}
If we define
\begin{equation}
\tilde q=q+\Delta=p_1-p\,
\end{equation}
It can be seen that the factorization prescription given by (\ref{eq:W_factored}) is the same as the deForrest cc2 prescription\cite{DeForest:1983vc} with modification for the covariant normalization of the Dirac spinors.

Factored response functions defined by $R_i=r_i n_+(p)$ can then be written as
\begin{align}
r_L=&\frac{1}{64\pi m_N^4}\{-4 F_1^2(Q^2) m_N^2 Q^2 - 8 F_1(Q^2) F_2(Q^2) m_N^2
  (\nu^2 + Q^2) + 4 E_p^2 (4 F_1^2(Q^2) m_N^2 \nonumber\\
   & +F_2^2(Q^2) Q^2)+ 4 E_p \nu (4 F_1^2(Q^2) m_N^2 +
   F_2^2(Q^2) Q^2) + F_2^2(Q^2) (\nu^2 Q^2 -
   4 m_N^2 (\nu^2 + Q^2))\nonumber\\
&-2\delta (2 E_p + \nu) (-4 F_1^2(Q^2) m_N^2 +
  F_2^2(Q^2) (2 E_p \nu + \nu^2 - Q^2))  \nonumber\\
& +\delta^2[-4 E_p^2 F_2^2(Q^2) + 4 F_1^2(Q^2) m_N^2 -
 12 E_p F_2^2(Q^2) \nu + F_2^2(Q^2) (-5 \nu^2 + Q^2)] \nonumber\\
&-4\delta^3 F_2^2(Q^2) (E_p + \nu)-\delta^4F_2^2(Q^2)\}\,,
\end{align}
\begin{align}
r_T=&\frac{1}{64\pi m_N^4}\{4 [4 F_1(Q^2) F_2(Q^2) m_N^2 Q^2 +
  F_2^2(Q^2) (2 m_N^2 + p_\perp^2) Q^2 +
  2 F_1^2(Q^2) m_N^2 (2 p_\perp^2 + Q^2)]\nonumber\\
  &-16\delta F_1(Q^2) (F_1(Q^2) + F_2(Q^2)) m_N^2 \nu+\delta^2(8 E_p^2 F_2^2(Q^2) - 8 F_1^2(Q^2) m_N^2 +
   8 E_p F_2^2(Q^2) \nu \nonumber\\
&- 2 F_2^2(Q^2) Q^2)+4\delta^3 F_2^2(Q^2) (2 E_p + \nu)+2\delta^4 F_2^2(Q^2)\}\,,
\end{align}
\begin{align}
r_{TT}=&\frac{-4 p_\perp^2 (4 F_1^2(Q^2) m_N^2 + F_2^2(Q^2) Q^2)}{64\pi m_N^4}
\end{align}
and
\begin{align}
r_{LT}=&\frac{1}{64\pi m_N^4}4 \sqrt{2}\{ (2 E_p + \nu) p_\perp
 (4 F_1^2(Q^2) m_N^2 + F_2^2(Q^2) Q^2)\nonumber\\
 &+\delta p_\perp [4 F_1^2(Q^2) m_N^2 +
   F_2^2(Q^2) (-2 E_p \nu - \nu^2 + Q^2)]-\delta^2 F_2^2(Q^2) \nu p_\perp
\}\,,
\end{align}
where $p_\perp$ is the magnitude of the component of $\bm{p}$ perpendicular to $\bm{q}$.

The reduction factor can then be written as
\begin{equation}
k \sigma_{ep} =  \frac{m_p \, m_n \, p_p}{8 \pi^3 \, M_d} \,
\sigma_{Mott} \,
f_{rec}^{-1} \,
 \left[  v_L r_L +   v_T r_T
 + v_{TT} r_{TT}\cos 2\phi_p + v_{LT}r_{LT}\cos\phi_p
\right] \,.
\end{equation}

\bibliography{mom_dist}

\end{document}